\documentclass[manuscript]{aastex} 

\usepackage[colorlinks=true,linkcolor = blue, anchorcolor = red, citecolor = blue, filecolor = red, pagecolor = red, urlcolor = red]{hyperref}
\usepackage{lscape,longtable} 
\usepackage{chngpage} 
\usepackage{color}
\usepackage{CJKfntef}

\newcommand{\lamost}{{\sc lamost}}
\newcommand{\kepler}{{\it Kepler}}
\newcommand{\kasc}{{\sc kasc}}
\newcommand{\lk}{{\sc lamost}-{\it Kepler}}

\newcommand{\teff}{$T_{\rm eff}$}
\newcommand{\teffL}{$T_{\rm eff, LASP}$}
\newcommand{\teffH}{$T_{\rm eff, Huber}$}

\newcommand{\logg}{$\log g$}
\newcommand{\loggL}{$\log g_{\rm LASP}$}
\newcommand{\loggH}{$\log g_{\rm Huber}$}

\newcommand{\vsini}{$v \sin i$}

\newcommand{\feh}{$\rm [Fe/H]$}
\newcommand{\fehL}{$\rm [Fe/H]$_{\rm LASP}}
\newcommand{\fehH}{$\rm [Fe/H]$_{\rm Huber}}

\newcommand{\lkfields}{LK-fields}
\newcommand{\project}{LK-project}
\newcommand{\kp}{$K\rm p$}

\newcommand{\snrg}{SNR$_g$}

\newcommand{\vrad}{$v_{\rm rad}$}

\newcommand{\ulyss}{{\sc ul}y{\sc ss}}
\newcommand{\cfi}{{\sc cfi}}
\newcommand{\lasp}{{\sc lasp}}
\newcommand{\kms}{km\,s$^{-1}$}

\shorttitle{\lamost\ observation in the \kepler\ Field.}
\shortauthors{Ren et al.}

\begin{document}

\title{\lamost\ observations in the \kepler\ field. Analysis of the stellar parameters measured with the \lasp\ based on the low-resolution spectra\altaffilmark{*}}

\author{
Anbing Ren\altaffilmark{1}, 
Jianning Fu\altaffilmark{1,**}, 
Peter De Cat\altaffilmark{2}, 
Yue Wu\altaffilmark{3}, 
Xiaohu Yang\altaffilmark{1},  
Jianrong Shi\altaffilmark{3}, 
Ali Luo\altaffilmark{3}, 
Haotong Zhang\altaffilmark{3},  
Subo Dong\altaffilmark{4}, 
Ruyuan Zhang\altaffilmark{1}, 
Yong Zhang\altaffilmark{5}, 
Yonghui Hou\altaffilmark{5}, 
Yuefei Wang\altaffilmark{5}, 
Zihuang Cao\altaffilmark{3} and 
Bing Du\altaffilmark{3}
}
\altaffiltext{1}{Department of Astronomy, Beijing Normal University, No. 19 Xinjiekouwai Street, Haidian District, Beijing 100875, China, renanbing@gmail.com}
\altaffiltext{2}{Royal observatory of Belgium, Ringlaan 3, B-1180 Brussel, Belgium}
\altaffiltext{3}{Key Lab for Optical Astronomy, National Astronomical Observatories, Chinese Academy of Sciences, Beijing 100012, China}
\altaffiltext{4}{Kavli Institute for Astronomy and Astrophysics, Peking University, Yi He Yuan Road 5, Hai Dian District, Beijing 100871, China}
\altaffiltext{5}{Nanjing Institute of Astronomical Optics \& Technology, National Astronomical Observatories, Chinese Academy of Sciences, Nanjing 210042, China}
\altaffiltext{*}{Based on observations collected with the Large Sky Area Multi-Object Fiber Spectroscopic Telescope (\lamost) located at the Xinglong Observatory, China.}
\altaffiltext{**}{Send offprint request to: jnfu@bnu.edu.cn}

\received{\underline{05-09-2016}}
\revised{\underline{06-19-2016}}
\accepted{\underline{06-27-2016}}

\begin{abstract}
All of the 14 subfields of the \kepler\ field have been observed at least once with the Large Sky Area Multi-Object Fiber Spectroscopic Telescope (\lamost, Xinglong Observatory, China) during the 2012-2014 observation seasons. There are 88,628 reduced spectra with \snrg\ (signal-to-noise ratio in $g$ band) $\geq$ 6 after the first round (2012-2014) of observations for the \lamost-\kepler\ project (\project). By adopting the upgraded version of the \lamost\ Stellar Parameter pipeline (\lasp), we have determined the atmospheric parameters (\teff, \logg, and \feh) and heliocentric radial velocity \vrad\ for 51,406 stars with 61,226 spectra. Compared with atmospheric parameters derived from both high-resolution spectroscopy and asteroseismology method for common stars in \citet{Hub2014}, an external calibration of \lasp\ atmospheric parameters was made, leading to the determination of external errors for the giants and dwarfs, respectively. Multiple spectroscopic observations for the same objects of the \project\ were used to estimate the internal uncertainties of the atmospheric parameters as a function of \snrg\ with the unbiased estimation method. The \lasp\ atmospheric parameters were calibrated based on both the external and internal uncertainties for the giants and dwarfs, respectively. A general statistical analysis of the stellar parameters leads to discovery of 106 candidate metal-poor stars, 9 candidate very metal-poor stars, and 18 candidate high-velocity stars. Fitting formulae were obtained segmentally for both the calibrated atmospheric parameters of the \project\ and the KIC parameters with the common stars. The calibrated atmospheric parameters and radial velocities of the \project\ will be useful for studying stars in the \kepler\ field.
\end{abstract}


\section{Introduction}
The main scientific objective of the NASA (National Aeronautics and Space Administration) space mission \kepler\ are to detect the Earth-size and even larger planets in the habitable zone \citep{Kas1993,Bor2007} of solar-like stars by using the method of photometric transits \citep{Bor2009}, and to determine the properties of the planet host stars by means of asteroseismic methods \citep{Chr2007}. Since the successful launch of \kepler\ on March 7, 2009, the number of stars having photometric time-series with an ultra-high precision of a few micro-magnitudes has increased steadily over a time-span of 4 years. As a large number of uninterrupted time-series has been obtained for pulsating stars of all kinds and flavors, the \kepler\ mission provides an unprecedented opportunity to study stellar oscillations. The \kepler\ Asteroseismic Science Consortium (\kasc, \citealt{Chr2007}), with a broad community participation, was established to select the most promising asteroseismic targets in the \kepler\ field of view (hereafter `\kepler\ field') as targets for \kepler\ and to study their internal structure by means of asteroseismic methods \citep{Gil2010,Cha2010}. However, a reliable asteroseismic modeling requires reliable basic stellar physical parameters such as atmospheric parameters (the effective temperature \teff, the surface gravity \logg, and the metallicity \feh) and the projected rotational velocity (\vsini). Unfortunately, the atmospheric parameters as given in the \kepler\ Input Catalogue (KIC, \citealt{Bro2011}) are not always unsuited for a successful asteroseismic modeling as their errors amount to $\sim$200 K in \teff, and to $\sim$0.5 dex in both \logg\ and \feh. Moreover, KIC atmospheric parameters are missing for a significant fraction of the \kepler\ objects. The shortcomings of the stellar properties in the KIC have been quantified in follow-up studies and are summarized by \citet[hereafter H14]{Hub2014}. 

During the last decade, a lot of ground-based observations have been gathered for a wide variety of \kepler\ targets to support their space-based observations. An enormous observational effort involving 2-m class telescopes located in 12 countries in the northern hemisphere has been coordinated by \citet{Uyt2010a,Uyt2010b} for the observations of \kasc\ objects, which leads to the characterization of, amongst others, OB-type stars \citep[including candidate $\beta$\,Cephei and slowly pulsating B stars;][]{Cat2010, Leh2011,Tka2013}, AF-type stars \citep[including candidate $\delta$\,Scuti and $\gamma$\,Doradus stars;][]{Cat2011,Tka2012,Tka2013,Nie2015}, solar-like stars \citep{Mol2008,Bru2012,Kar2013}, giants \citep{Bru2011}, and red giants \citep{Thy2012}. Though strong efforts have been made to characterize all types of asteroseismic targets, a significant fraction of the \kasc\ targets remained unobserved, mainly because of the faintness of the targets and the unavailability of a sufficient amount of telescope time. 

H14 presented an improved catalog for 196,468 stars observed by the NASA \kepler\ mission to support the study of the planet-occurrence rate by consolidating the published values of the atmospheric parameters (\teff, \logg, \feh) that are derived with different observational techniques (mainly photometry, spectroscopy, asteroseismology and exoplanet transits). 
It is a valuable contribution to the improvement of the stellar properties of \kepler\ targets, but for a considerable fraction of stars, the KIC parameters could not be updated. Moreover, the consistency in the results is lacking as they are based on observations from heterogeneous devices and analysis techniques.

The \lamost\ (the Large Sky Area Multi-Object Fiber Spectroscopic Telescope, also called the GuoShouJing Telescope) \citep{Su1998,Zhao2012}, is a special 4-meter reflecting Schmidt telescope located at the Xinglong station of the National Astronomical Observatories of China \citep{Cui2012,Luo2012}. Its focal length is 20 m and the focal plane, with a diameter of 1.75 m corresponding to a circular field of view of 5 degrees on the sky \citep{Wang1996}, is covered with 4000 optical fibres connected to 16 two-arm low-resolution spectrographs with 32 CCD cameras. \lamost\ spectra have a resolution of about 1800 and cover the wavelength range 370-900 nm \citep{Cui2012,Zhao2012}. The combination of a large aperture and a wide field of view covered by 4000 fibers makes \lamost\ the most powerful optical spectroscopic survey instrument in the northern hemisphere at present. We therefore initiated the \lk\ project \citep[\project;][]{Dec2014} to acquire \lamost\ spectra for as many objects in the \kepler\ field as possible and to characterize them in terms of spectral classification (spectral type with any peculiarities), atmospheric parameters (\teff, \logg, \feh), rotation rate (\vsini) and radial velocity (\vrad).
It is the only way we can derive these parameters with an accuracy required for a detailed asteroseismic study for the vast majority of the \kepler\ objects in an efficient and homogeneous way. For a detailed description of the \project, we refer interested readers to \citet{Dec2015}.

The low-resolution \lamost\ spectra available in the catalogue of the \project\ \citep{Dec2015} have been analysed by three different teams, each with their own independent method. This paper presents the analysis of the released parameters from the `Asian team' (composed by the \lamost\ project's data processing department group and ABR, JNF, XHY group from Beijing Normal University) who determined the stellar atmospheric parameters by using the official \lamost\ Stellar Parameter pipeline \citep[\lasp,][]{Wu2014,Wu2011a,Luo2015}. This paper is organized as follows. In Sections \ref{sect:2} and \ref{sect:3}, brief descriptions of the observations and spectral data are given, respectively. A concise introduction of the \lasp\ stellar parameter calculation is given in Section \ref{sect:4}. In Section \ref{sect:5}, \lasp\ stellar parameters and their errors are calibrated based on the results from both external calibrations and internal uncertainties for the giants and dwarfs, respectively. 
Then we perform a statistical analysis of the calibrated atmospheric parameters and radial velocities present in the \lasp\ catalogue in Section \ref{sect:6}. It includes identifications of candidates of particular objects. 
We compare calibrated \lasp\ parameters with the values in KIC for common stars in Section \ref{sect:7}. The paper is ended with conclusions and the prospect of the \project\ in Section \ref{sect:8}.

\section{Observations} \label{sect:2}
\subsection{Observation Plan} \label{sect:2.1}
The \kepler\ field is relatively large (105 deg$^{2}$). Fourteen circular \lamost-\kepler\ fields (\lkfields) with a diameter of 5 degrees are needed for a close-to-full coverage of the \kepler\ field \citep[see Figure 2 of][]{Dec2015}. Some of these \lkfields\ are overlapped. To prepare the \lamost\ observations of the 14 requested \lkfields, we constructed a prioritized target list which consists of targets from the KIC (`\kepler\ targets') supplemented with objects that have an absolute magnitude less than 20 in V band from the USNO-B catalogue \citep[`field targets';][]{Mon2003}. We prioritized these objects within the \kepler\ field based on their coordinates (R.A. (2000) and Dec. (2000)), their brightness (the KIC magnitude \kp\ for most of the objects), the  availability of their stellar parameters in the KIC (the effective temperature \teff, the surface gravity \logg, and the metallicity \feh), and their scientific importance within the research community involved in the Kepler \citep[see Section 2 of][]{Dec2015}. For the first round of observations (2012-2014), from which the resulting \lamost\ spectra are analysed in this paper, the top priority was assigned to the $\sim$6500 targets that were of scientific interest for the \kepler\ Asteroseismic Science Consortium (\kasc) at the beginning of the \project. Note that the \kepler\ field contains four open clusters, namely NGC6791, NGC6811, NGC6819 and NGC6866. Due to the density of the stars in the \kepler\ field, especially the regions containing the open clusters, it is impossible to observe all \kepler\ targets with \lamost\ with only one round of observations. However, now that the observations of the \kepler\ mission have ended, the priority has shifted in the next round of observation to the targets for which \kepler\ observations are available but for which no high-quality \lamost\ spectrum is available yet. For more details about the construction of the target list, we refer the interested reader to Section 2 of \citet{Dec2015}.

\subsection{Observation Progress}
\label{sect:2.2}
The observations for the \project\ started in May 2011, during the pilot survey period of \lamost. As the observations started in the test phase of the \lamost, the observations suffered from hardware and software failures in the beginning of the \project. Moreover, the \kepler\ field is best visible in the northern summer during which \lamost\ is closed for several months due to the Monsoon. Therefore, some of the \lkfields\ had to be observed several times, and four observation seasons of the \kepler\ field were needed to complete the first round of observations (May 2011 - September 2014) in which all \lkfields\ were observed at least once under good conditions. Note that the spectra observed during the test phase of the \lamost\ (before October 24, 2011) were omitted from the data products of the latest version (V2.7.5) of the data reduction and analysis pipeline because the telescope and instruments were still in the debugging stage at that time. The instability of the whole system led directly to the relatively low quality of the spectra. Although the catalogue of the \project\ contains a small proportion of spectra taken during the pilot survey with a sufficient quality, they are kept out of the catalogue of the \lamost\ official data release. The updated \lasp\ will not be applied to determine stellar parameters from the spectra that are observed before October 2011 and hence will not be included in the official data release of the \lamost. So these spectra are omitted from our study too. In total, the 14 \lkfields\ were observed with 35 plates during 25 nights in the July 2012 - September 2014 observations seasons:
\\
$\bullet$ 2012: 3 \lkfields\ with   7 plates during   3 nights, \\
$\bullet$ 2013: 6 \lkfields\ with 14 plates during 12 nights, \\
$\bullet$ 2014: 7 \lkfields\ with 14 plates during 10 nights. \\

For a detailed overview of the progress of the observations within the \project, we refer interested readers to Tables 1 \& 2 and Section 5 of \citet{Dec2015}. In Table \ref{Tab1}, we give an overview of the number of results obtained by the Asian team based on the \lamost\ spectra gathered up to the end of the 2014 observation season of the \kepler\ field.

\clearpage
\begin{table} \scriptsize 
\tabletypesize{\scriptsize}
\begin{center}
\caption[]{}{The progress of observation and achieved parameters during the 2012-2014 observations seasons for the LK-project. This table follows the format of the Table 1 in De Cat et al. (2015). In the upper part of the table, we list the name of the LK-field (LK-field), right ascension (R.A.(2000)) and Declination (Dec.(2000)) of the central bright star, the open cluster name in the LK-field (Cluster), the  date of observation (Date), the number of plates that have been obtained for each LK-field (\#), the total number of spectra (Spectra), the total number of stellar parameters (Parameters) resulting from the \lasp, and the number of the objects which were observed photometrically by the \kepler\ mission (KO). In the bottom part of the table, we give for each category the total number of analysed spectra (Total), the number of different targets (Unique), and the number of targets that have been observed from one time to at least five times (1x, 2x, 3x, 4x, and +5x, respectively). \label{Tab1}}\\ 
\begin{tabular}{lllcccrrr}
\hline\noalign{\smallskip}
\tableline
LK-field & R.A.(2000)     & Dec.(2000)      & Cluster      & Date           &\# & Spectra & Parameters & KO  \\
\hline\noalign{\smallskip}                                                 
LK01     & 19:03:39.258 & +39:54:39.24 &                  & 2014/06/02 & 2 &  4944    &  3481          & 1851\\
LK02     & 19:36:37.977 & +44:41:41.77 & NGC6811 & 2012/06/04 & 1 &   506     &   315           &   195\\
              &                       &                        &                  & 2014/09/13 & 2 &  6365    & 4903           & 3166\\
LK03     & 19:24:09.919 & +39:12:42.00 & NGC6791 & 2012/06/15 & 3 &  8490    & 6085           & 4169\\
LK04     & 19:37:09.862 & +40:12:49.63 & NGC6819 & 2012/06/17 & 3 &  7612    & 4172           & 2861\\
LK05     & 19:49:18.139 & +41:34:56.85 &                  & 2013/10/05 & 2 &  5744    & 3845           & 2346\\
               &                      &                        &                  & 2014/05/22 & 1 &  2336    &   883           &   683\\
LK06     & 19:40:45.383 & +48:30:45.10 &                  & 2013/05/22 & 1 &  2480    & 1486           & 1145\\
              &                       &                        &                  & 2013/05/23 & 1 &  1989    &   798           &   670\\
              &                       &                        &                  & 2013/09/14 & 1 &  2745    & 2212           & 1543\\
LK07     & 19:21:02.816 & +42:41:13.07 &                  & 2013/05/19 & 1 &  3136    & 2160           & 1652\\
              &                       &                        &                  & 2013/09/26 & 1 &  2922    & 2412           & 1818\\
LK08     & 19:59:20.425 & +45:46:21.15 & NGC6866 & 2013/09/25 & 2 &  5464    & 4079           & 1757\\
              &                       &                        &                  & 2013/10/02 & 1 &  2494    &   436           &       5\\
              &                       &                        &                  & 2013/10/17 & 1 &  2427    & 1286           &   617\\
              &                       &                        &                  & 2013/10/25 & 1 &  2708    & 2057           &   827\\
LK09     & 19:08:08.340 & +44:02:10.88 &                  & 2013/10/04 & 1 &  2856    & 2387           & 1618\\
LK10     & 19:23:14.829 & +47:11:44.80 &                  & 2014/05/20 & 2 &  2785    & 1802           & 1239\\
LK11     & 19:06:51.499 & +48:55:31.77 &                  & 2014/09/18 & 1 &  2852    & 2563           & 1619\\
LK12     & 18:50:31.041 & +42:54:43.72 &                  & 2013/10/07 & 1 &  2643    & 2347           & 1284\\
LK13     & 18:51:11.993 & +46:44:17.52 &                  & 2014/05/02 & 1 &  2548    & 1917           & 1074\\
              &                       &                        &                  & 2014/05/29 & 2 &  4697    & 3553           & 1901\\
LK14     & 19:23:23.787 & +50:16:16.64 &                  & 2014/09/17 & 1 &  2821    & 2605           & 1391\\
              &                       &                        &                  & 2014/09/27 & 1 &  2457    & 1578           &   803\\
              &                       &                        &                  & 2014/09/29 & 1 &  2607    & 1864           &   951\\
\hline\noalign{\smallskip}                                                             
Total      &                       &                        &                  &                    &35 & 88628  & 61226         & 37185\\
Unique  &                       &                        &                  &                    &     &             & 51406         & 30110\\
1x          &                       &                        &                  &                    &     &             & 42773         & 23950\\
2x          &                       &                        &                  &                    &     &             &   7550         &   5325\\
3x          &                       &                        &                  &                    &     &             &     986         &     762\\
4x          &                       &                        &                  &                    &     &             &       92         &       68\\
+5x        &                       &                        &                  &                    &     &             &         5         &         5\\
\noalign{\smallskip}\hline
\end{tabular}
\end{center}
\end{table}
 
\section{Spectral Data}
\label{sect:3}
\subsection{Data Reduction}
\label{sect:3.1}
\lamost\ has an automatic software system for its observation and data processing procedure \citep{Luo2004,Wu2014}. The goals of this system are to classify spectra and to calculate parameters from these spectra. 
The version V2.7.5 of the spectral reduction and analysis pipeline are used for the spectra obtained in the period 2012.06-2014.09. CCD Raw images from the \project\ were reduced and analyzed by the standard \lamost\ automated data reduction and analysis system including the 2-dimension (2D) reduction pipeline \citep{Luo2004}, the 1-dimension (1D) pipeline, and the \lamost\ stellar parameter pipeline - \lasp\ \citep[]{Wu2014,Luo2015}. The CCD images are fed to the 2D reduction pipeline, which conforms with the spectral 2D pipeline of SDSS \citep{Sto2002}, to extract calibrated 1D spectrum in the format of FITS file for each object. The 2D pipeline includes seven basic tasks: dark and bias subtraction, flat field correction, spectra extraction, sky subtraction, wavelength calibration, sub-exposure merging and wavelength band combination. Simultaneously, the uncertainties in the wavelength and relative flux calibration are calculated in detail during this reduction process \citep{Luo2015}. The 1D pipeline, which is based on the specBS pipeline used for the analysis of SDSS spectra \citep{Gla1998}, is to determine the spectral type of the stars and to automatically measure either the \vrad\ for stars or the redshift for galaxies and quasi-stellar objects (QSOs) by template matching and using a line recognition algorithm. The stellar templates were constructed by using accumulated \lamost\ DR1 spectra \citep{Wei2014}. The main output of the spectral processing includes calibrated spectra with the corresponding analysis results and a catalog with information about the processed objects.

\subsection{Data Release}
\label{sect:3.2} 
\begin{figure} 
\epsscale{0.80}
\plotone{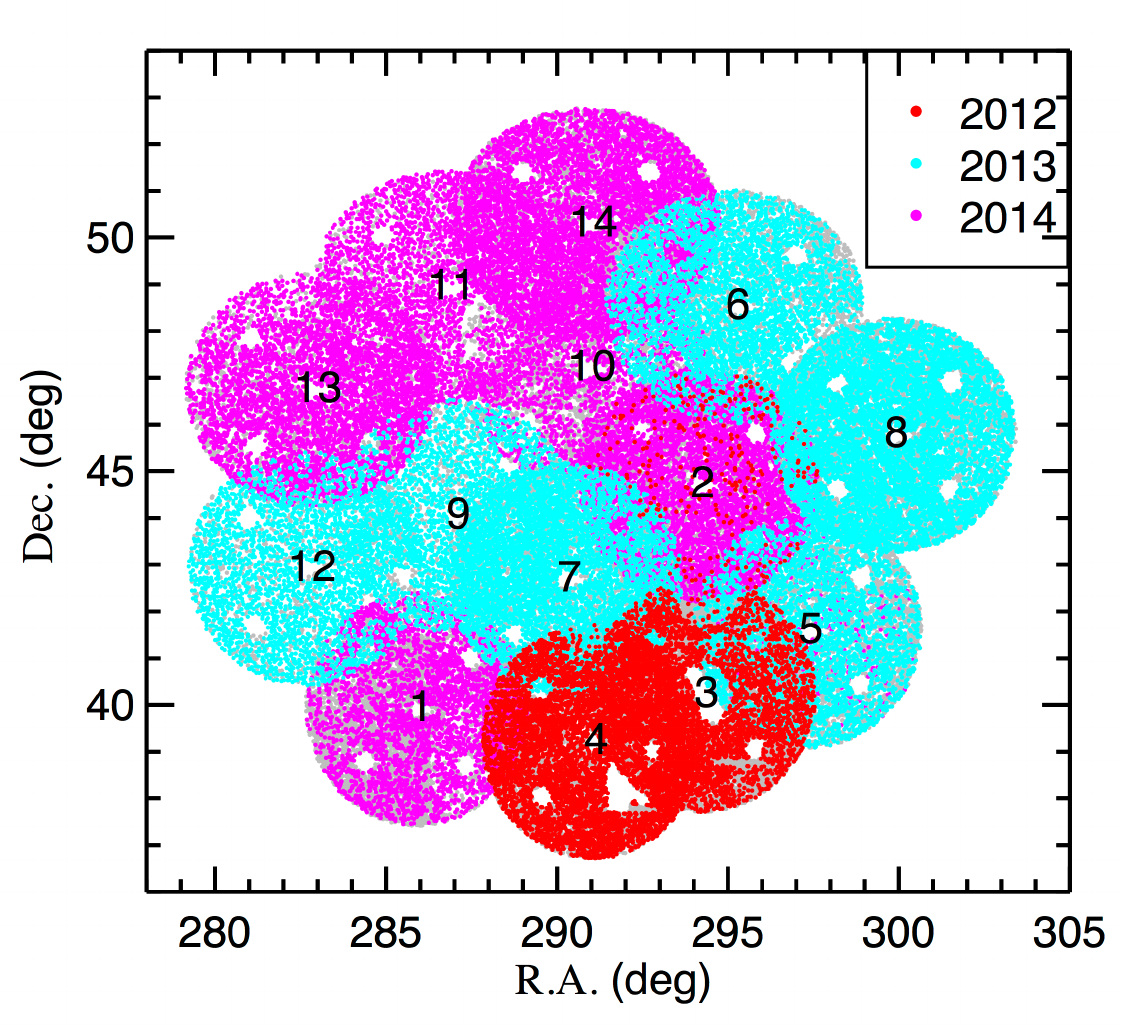}
\caption{The spatial distribution of all the targets that were observed during the 2012-2014 observation seasons of the \project. The 51,406 targets for which we could derive the stellar parameters with \lasp\ are given in colour (red for 2012; cyan for 2013; magenta for 2014) while all the others are given in grey. }
\label{Fig1} 
\end{figure}

The updated and calibrated spectra are provided to the astronomers at regular intervals. Up to September 2014, a total of 61,226 flux- and wavelength-calibrated, sky-subtracted low-resolution ($R$=1800) spectra and their \lasp\ stellar parameters were obtained during the first round of observations for the \project.  These data will be released to the public along with the third data release (DR3\footnote{http://dr3.lamost.org}) in June 2017. They can be downloaded from the official \lamost\ website\footnote{www.lamost.org}. The spatial distribution of the targets observed during the 2012-2014 observation seasons for the \project\ is shown in Figure \ref{Fig1}. The database of low-resolution spectra of the \project\ consists of \kepler\ (61,218) and field targets (8) nearly covering the whole region of the \kepler\ field, except for one center and 4 off-center circle holes in each plate which contain the central bright star ($V < 8$; for the adaptive optics wave front sensor) and 4 guide stars ($V < 17$; for the guiding of the CCD cameras), respectively. 

\begin{figure} 
\epsscale{0.80}
\plotone{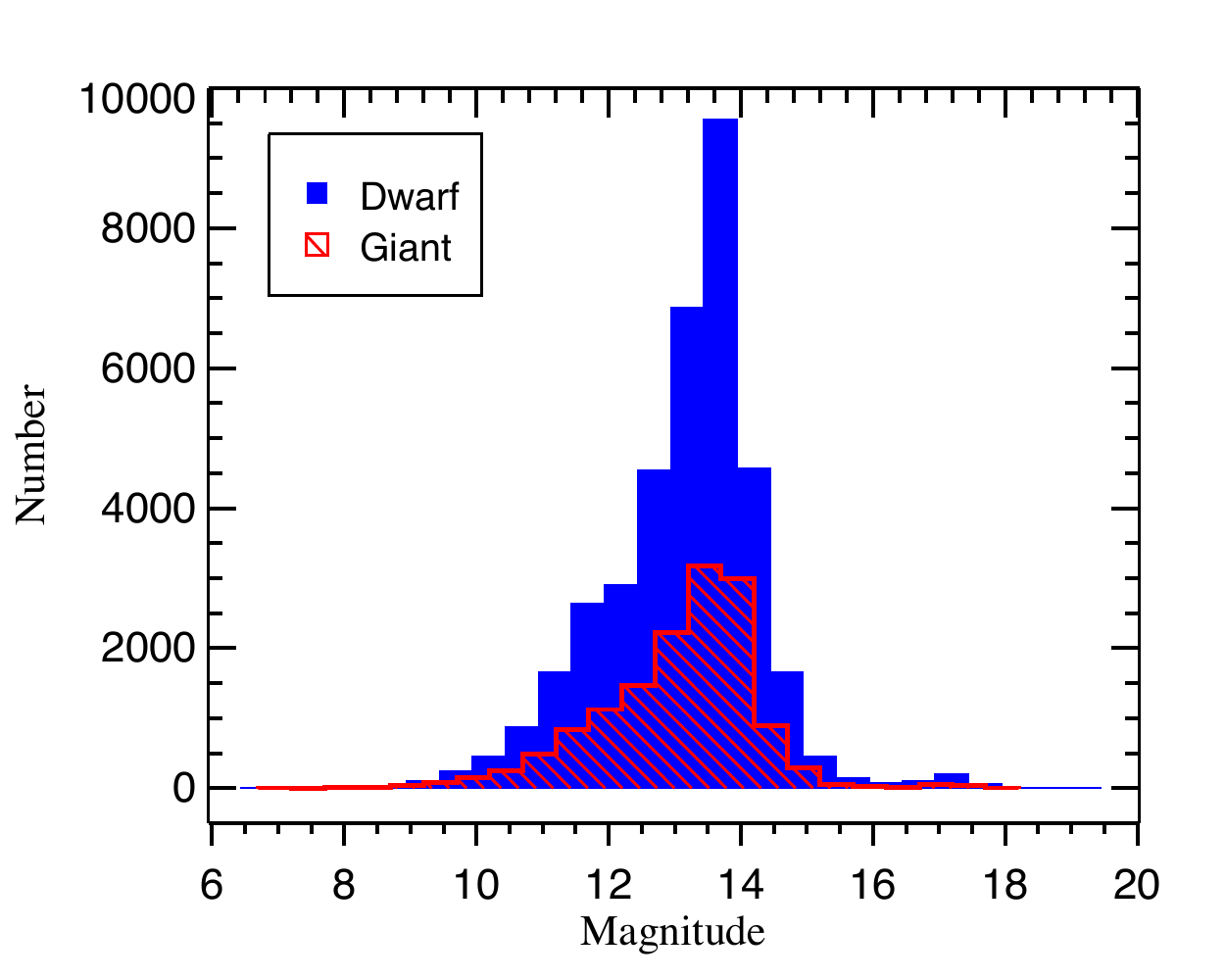} 
\caption{The distribution of the \kepler\ magnitude (\kp) as given in the KIC for the dwarfs (blue) and giants (red), respectively. \label{Fig2}}
\end{figure}

\begin{figure} 
\epsscale{0.80}
\plotone{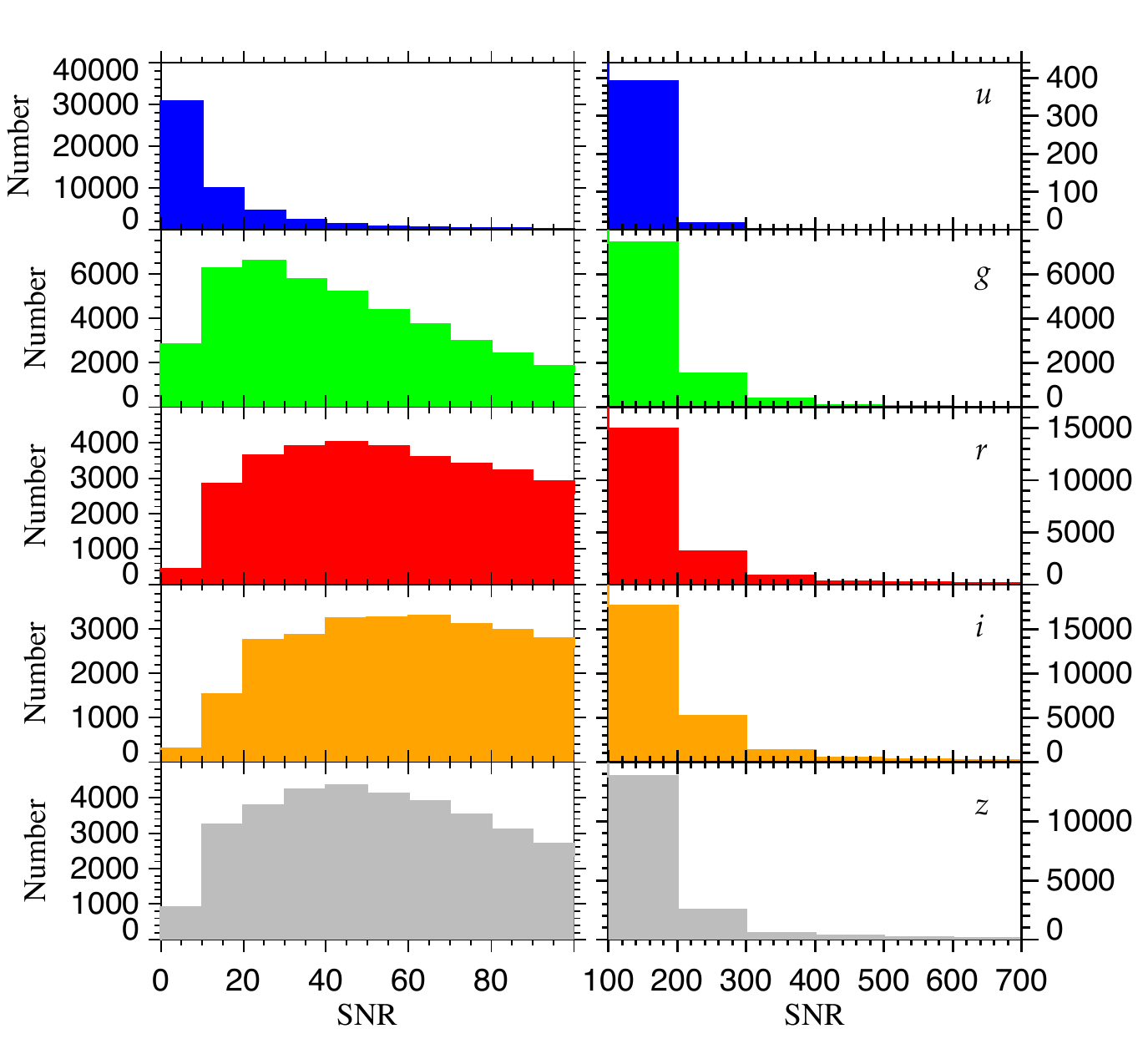} 
\caption{The distribution of the signal-to-noise ratio in the SDSS $u$, $g$, $r$, $i$ and $z$ bands, given from top to bottom respectively. The left and right panels show the SNR range from 0 to 100 with binsize 10 and the SNR range from 100 to 700 with binsize 100, respectively.  \label{Fig3}}
\end{figure}

Figure \ref{Fig2} displays the histogram of the \kepler magnitude distribution of the dwarfs and giants in the \project. These targets are mainly distributed in the range of 11-15 \kepler\ magnitude (\kp) which indicates that only a small portion of bright and faint targets have been observed in the first observation round of the \project.  The distribution of \kp\ also reflects the observation strategy of the \project\ under different conditions of observation. Indeed, the observations mainly focused on the very bright plates ($9 < r \leq 14$) in order to make full use of bright nights or nights with unfavourable weather conditions (e.g., poor seeing or low atmospheric transparency) \citep{Dec2015}. Note that the \kp\ are not given in the KIC for 22 objects.
 Figure \ref{Fig3} shows the histograms of the signal-to-noise ratio (SNR) in the Sloan Digital Sky Survey (SDSS) $u$, $g$, $r$, $i$ and $z$ bands for the spectra in the catalog of data release, from top to bottom respectively. We take \snrg\ $\geq$ 6 and \snrg\ $\geq$ 15 as the criterion to retain the \lamost\ spectra of the data release for stars that are observed during dark nights (eight nights before and after the new moon) and bright nights (all other nights except the three nights around the full moon in a lunar cycle), respectively \citep{Luo2015}. 
Hence, the distributions shown in the left panels of Figure \ref{Fig3} do not contain stars with \snrg\ $<$ 6 in the first bin (0-10).

Note that several problems have been found and fixed since the last DR2 update as described on the \lamost\ data release website\footnote{http://dr2.lamost.org}. Based on the updated information of the wrong IDs of fiber, 9 fibers' IDs of spectrograph 4 were found to be wrong since the survey in June 2012 for the observation of the \project. The regulation for calibrating the fibre position of spectrograph 4 is given in Table \ref{Tab2}. It includes the fibre unit ID on the focal plane (Unitid) as given in the headers of the \lamost\ 1D fits spectra, the old ID of the fibre (Fbid\_old) and the corrected ID of the fibre (Fbid\_now). Since the last three digits in the name of the \lamost\ 1D fits file (Filename) represent the fibre's ID, the wrong IDs of fibres lead to 71 Filenames in Table 4 of \citet{Dec2015} are different from the Filenames in our Table \label{Tab3}. We adjusted the 71 Filenames following the rules of calibration for the fibre position in order to obtain the KIC IDs for observed targets from the Table 4 by matching the Filenames. The revised Filenames are indicated with a letter `c' after the name of the fits file in the `Filename' column of Table \ref{Tab3}. 
\begin{deluxetable}{lrr} 
\tabletypesize{\scriptsize}
\tablecaption{The regulation for calibrating the fibre position of spectrograph 4. \label{Tab2}}
\tablewidth{0pt}
\tablehead{ \colhead{Unitid} & \colhead{Fbid\_old} & \colhead{Fbid\_now} }
\startdata                                                                                                     
                                   F3229   &                            76   &                            87 \\
                                   F3230   &                            87   &                            79  \\
                                   F3231   &                            79   &                            95  \\
                                   F3232   &                            95   &                            84  \\
                                   F3333   &                            84   &                            76  \\
                                   H3129  &                            44   &                            31  \\
                                   H3130  &                            31   &                            46  \\
                                   H3131  &                            46   &                            26  \\
                                   H3132  &                            26   &                            44  \\
\enddata
\end{deluxetable}

\section{Stellar Parameter Calculation}
\label{sect:4}
An upgraded version of the \lasp\ was used to automatically determine the stellar atmospheric parameters (\teff, \logg\ and \feh) and radial velocity (\vrad) from fits spectra products of the 2D and 1D pipelines in the spectral database of the \project. The \snrg\ of these spectra are greater than or equal to 6 and 15 for the dark nights and bright nights, respectively. We select the spectra with STAR as $final\_class$ and late A, F, G or K as $final\_subclass$ to determine their stellar parameters. Both the \cfi\ (Correlation Function Initial) and \ulyss\ (Universit\'e de Lyon Spectroscopic analysis Software; \citealt{Wu2011b}) methods are consecutively applied to determine the stellar parameters in the processing procedure of \lasp. More information about the progress of stellar parameters calculation is given by \citet{Luo2015}.

We determined the stellar atmospheric parameters (\teff, \logg\, \feh), radial velocity (\vrad) and their errors for the selected 61,226 ($\sim$60.6\%) of the 101,086 low-resolution \lamost\ spectra available in the catalogue of the \project\ after completion of the first round of observations \citep{Dec2015}. As some of the \lkfields\ have been observed more than once and there are some overlaps between some of the fields, a fraction of the observed stars have multiple \lamost\ observations. Hence, the 61,226 analysed \lamost\ spectra correspond to 51,406 unique targets, including 671 A-type stars, 18,937 F-type stars, 25,847 G-type stars, and 5,952 K-type stars. These targets are indicated in colour on Figure \ref{Fig1}: red for 2012, cyan for 2013, and magenta for 2014. Apart from the central and off-center holes typical for each plate, it is clear that there are still several other block-like regions for which we could not determine the stellar parameters from the available \lamost\ spectra (given in grey). This is a reflection of problems with some spectrographs in these parts. 

In Table \ref{Tab3}, we give the information about these targets including the following columns:
\begin{enumerate}
\item column 1: The unique spectra ID (Obsid). 
\item column 2: The file name of the \lamost\ 1D fits file (Filename).
\item column 3: The final identification of the target after cross-identification with the Table 4 of \citet{Dec2015} by the Filename (Target). 
\item column 4: The observed right ascension in degrees (R.A.).
\item column 5: The observed declination in degrees (Dec.).
\item column 6: The magnitude in the KIC for this target (\kp).
\item column 7: The date and time of observation (yyyy-mm-ddThh:mm:ss.ss).
\item column 8: The spectral sub-class retrieved from the 1D fits files (Subclass).
\item column 9: The value of SNR in the $g$-band (\snrg).
\item column 10: the effective temperature (\teff) in K and its error giving by the \lasp.
\item column 11: The surface gravity (\logg) in dex and its error giving by the \lasp.
\item column 12: The metallicity (\feh) in dex and its error giving by the \lasp.
\item column 13: The radial velocity (\vrad) in \kms\ and its error giving by the \lasp.
\item column 14: The information whether the object has been observed by the \kepler\ mission (KO).
\end{enumerate}

\begin{deluxetable}{lllllll}  
\tabletypesize{\scriptsize} 
\rotate
\tablewidth{0pt} 
\tablecaption{The catalogue of the \lasp\ stellar parameters for the \project. \label{Tab3}} 
\tablehead{\colhead{Obsid} & \colhead{Filename} & \colhead{Target}  & \colhead{R.A. (deg)}   & \colhead{Dec. (deg)} & \colhead{\kp} &  \colhead{yyyy-mm-ddThh:mm:ss.ss} \\
\colhead{Subclass}  &  \colhead{\snrg}  & \colhead{\teff\ (K)}  & \colhead{\logg\ (dex)} & \colhead{\feh\ (dex)} & \colhead{\vrad\ ($km s^{-1}$)}              & \colhead{KO}   \\}
\startdata
    52201011  & spec-56083-IF04\_B56083\_sp01-011          & KIC07042868   & 294.51886  & 42.549595 &  9.117 & 2012-06-04T18:35:33.32 \\
   G5 & 76.73 & 4852.00$\pm$51.72  & 2.658$\pm$0.470 & -0.021$\pm$0.082 &   1.98$\pm$12.52 & N \\     
  52201018  & spec-56083-IF04\_B56083\_sp01-018          & KIC06957157   & 294.34372  & 42.407665 & 10.106 & 2012-06-04T18:35:33.32 \\
   G5 & 50.72 & 4742.09$\pm$64.08  & 2.497$\pm$0.455 & -0.002$\pm$0.098 &   6.07$\pm$11.74 & Y \\     
  \ldots    & \ldots    & \ldots     & \ldots      & \ldots      & \ldots      & \ldots \\
  243104076 & spec-56811-KP190339N395439V02\_sp04-076c  & KIC04347646   & 285.90503  & 39.466915 & 12.552 & 2014-06-02T18:57:40.48 \\
   F9 & 55.97 & 4941.71$\pm$126.43 & 2.355$\pm$0.768 & -0.694$\pm$0.208 & -14.05$\pm$19.06 & Y \\     
  243104077 & spec-56811-KP190339N395439V02\_sp04-077   & KIC04244973   & 286.03354  & 39.327650 & 10.305 & 2014-06-02T18:57:40.48 \\
   G5 & 20.67 & 4811.94$\pm$142.05 & 2.872$\pm$0.533 & -0.010$\pm$0.189 & -20.98$\pm$13.93 & Y \\     
  \ldots    & \ldots    & \ldots     & \ldots      & \ldots      & \ldots      & \ldots \\
  250013110 & spec-56930-KP192323N501616V03\_sp13-110   & 1350-10735728 & 293.58322  & 51.240780 &        & 2014-09-29T12:29:21.91 \\
   F6 & 16.53 & 6058.80$\pm$93.06  & 4.220$\pm$0.302 & -0.142$\pm$0.140 &  -9.03$\pm$16.17 & N \\     
  250013119 & spec-56930-KP192323N501616V03\_sp13-119   & KIC12361114   & 293.69366  & 51.197147 & 11.178 & 2014-09-29T12:29:21.91 \\
   F0 & 86.22 & 6708.83$\pm$51.12  & 3.536$\pm$0.265 & -0.012$\pm$0.088 & -35.16$\pm$17.78 & N \\     
  \ldots    & \ldots    & \ldots     & \ldots      & \ldots      & \ldots      & \ldots \\
  250016248 & spec-56930-KP192323N501616V03\_sp16-248   & KIC12933571   & 289.32538  & 52.398720 & 13.307 & 2014-09-29T12:29:22.02 \\
   K5 & 19.47 & 4480.25$\pm$49.50  & 2.678$\pm$0.333 & -0.266$\pm$0.077 & -89.24$\pm$8.66  & N \\     
  250016249 & spec-56930-KP192323N501616V03\_sp16-249   & KIC12883443   & 289.20694  & 52.250880 & 13.460 & 2014-09-29T12:29:22.02 \\
   K0 & 18.22 & 5286.27$\pm$102.40 & 3.808$\pm$0.468 &  0.035$\pm$0.138 & -49.52$\pm$13.55 & N \\     
\enddata
\flushleft The unique spectra ID (Obsid) is in the first column, the name of the \lamost\ 1D fits file (Filename) is in column 2, the KIC ID for these targets (Target) is in column 3,  the columns from 4 to 8 are for the information extracted from the headers of the \lamost\ 1D fits files: the observed right ascension (R.A.) and declination (Dec.) in degrees, the magnitude in the KIC (\kp), the mid-time of observation (yyyy-mm-ddThh:mm:ss.ss), the sub-classfication for targets (Subclass), the value of \snrg. The next four columns give the \lasp\ stellar parameters (\teff, \logg, \feh\ and \vrad) and their uncertainties. The last column gives the information whether the object has been observed by the \kepler\ mission (KO, Y=YES, N=NO). 
\end{deluxetable}

\section{Calibration of Stellar Parameters}
\label{sect:5}
The stellar atmospheric parameters of the \project\ targets that are listed in Table \ref{Tab3} are compared with the common targets in a sub-sample of H14 to perform an external calibration of \lasp\ atmospheric parameter uncertainties for the giants and dwarfs, respectively. On the other hand, multiple observation targets in the parameter catalogue of the \project\ are used to obtain the internal calibration of these parameter uncertainties. Then we recalibrate the \lasp\ stellar parameters based on the results of external calibration and internal calibration for the giants and dwarfs, respectively. 

\subsection{External Calibration} 
\label{sect:5.1}
\begin{figure}   
\centering
   \includegraphics[width=0.333\textwidth]{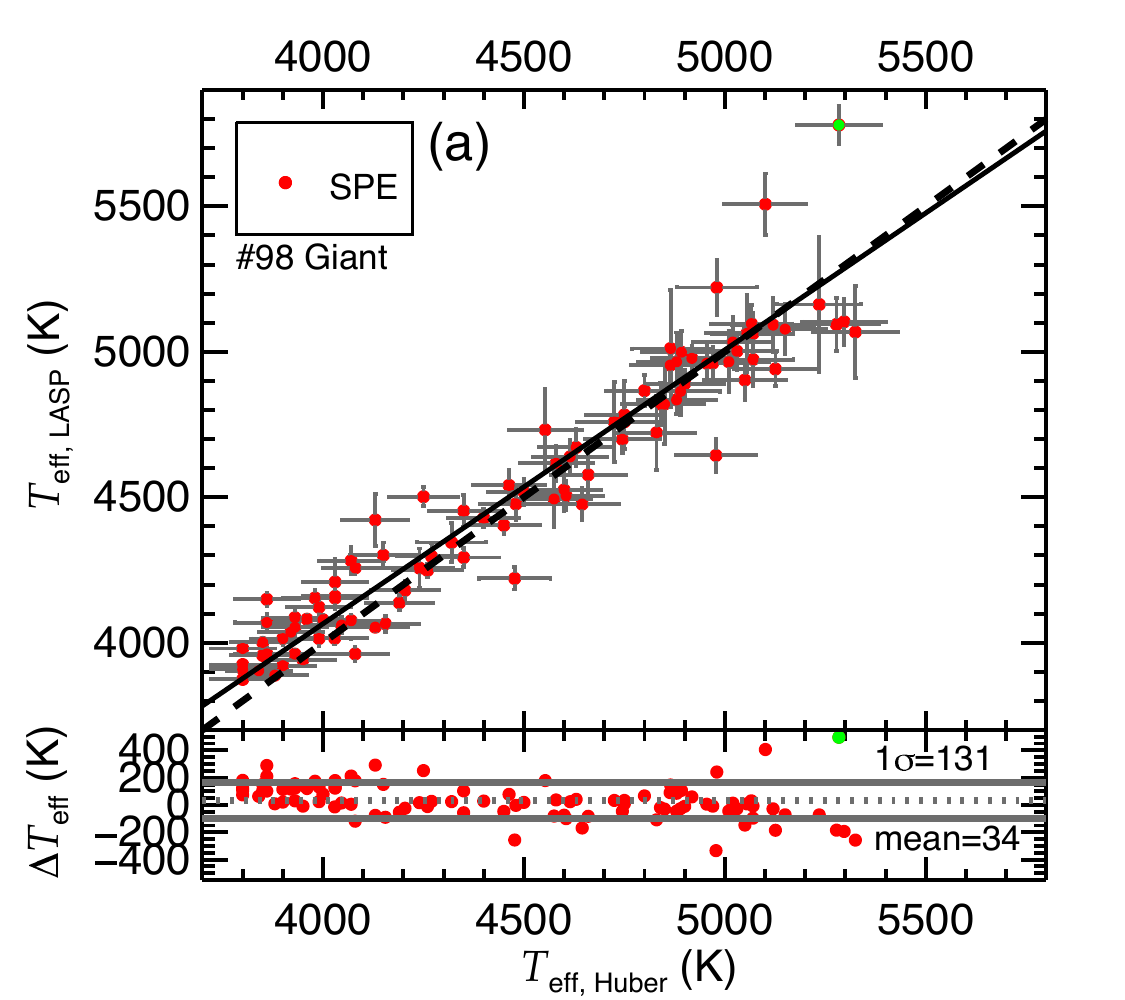}\hfill
   \includegraphics[width=0.333\textwidth]{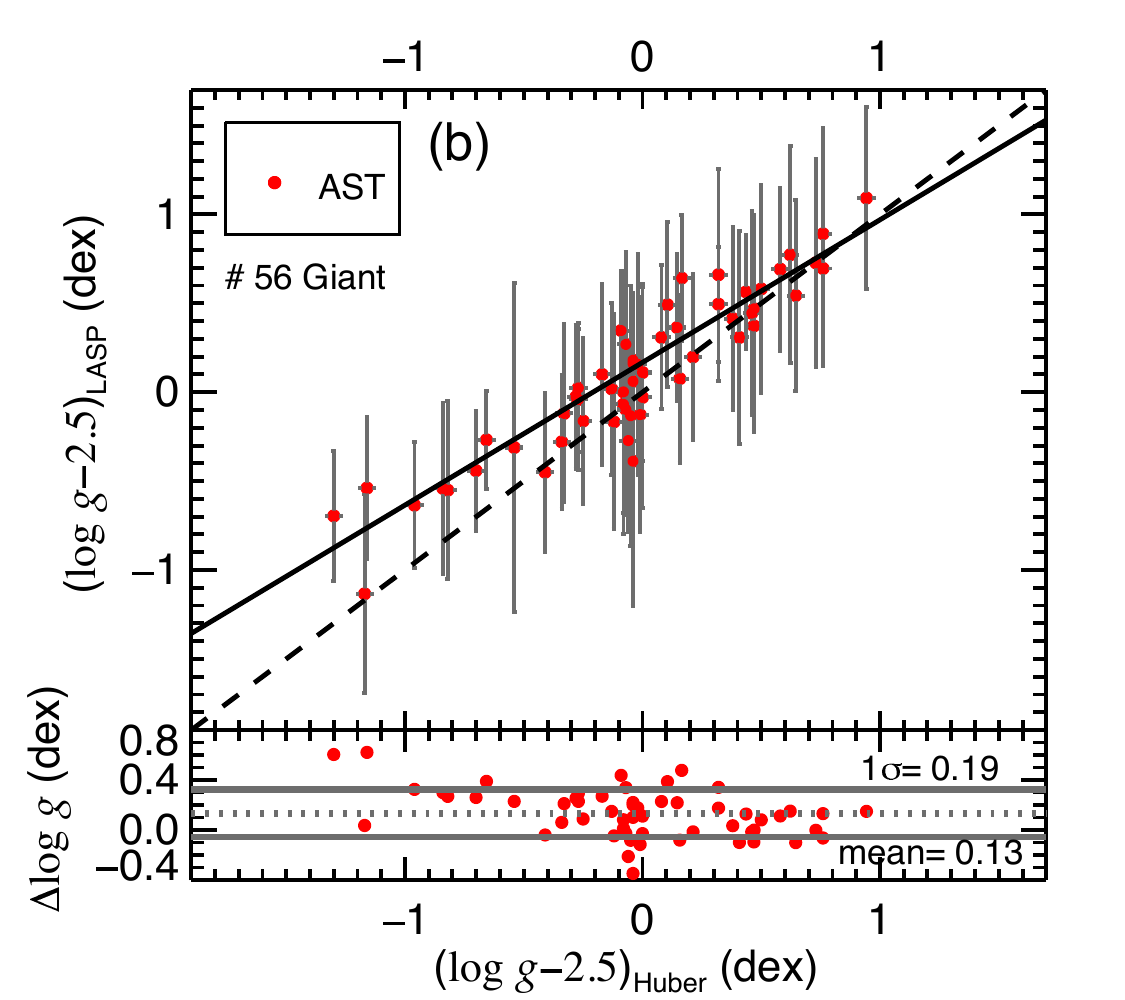}\hfill
   \includegraphics[width=0.333\textwidth]{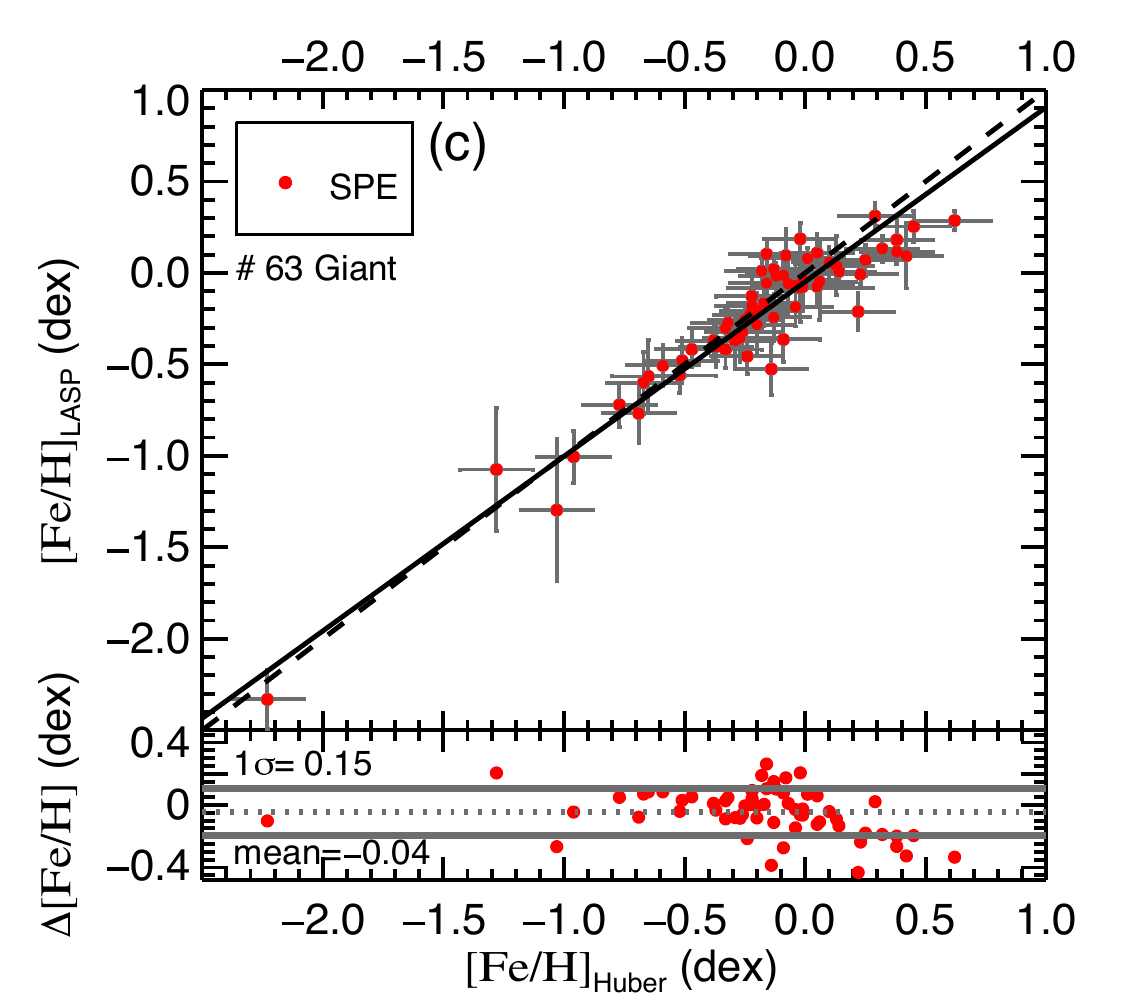}\\
   \includegraphics[width=0.333\textwidth]{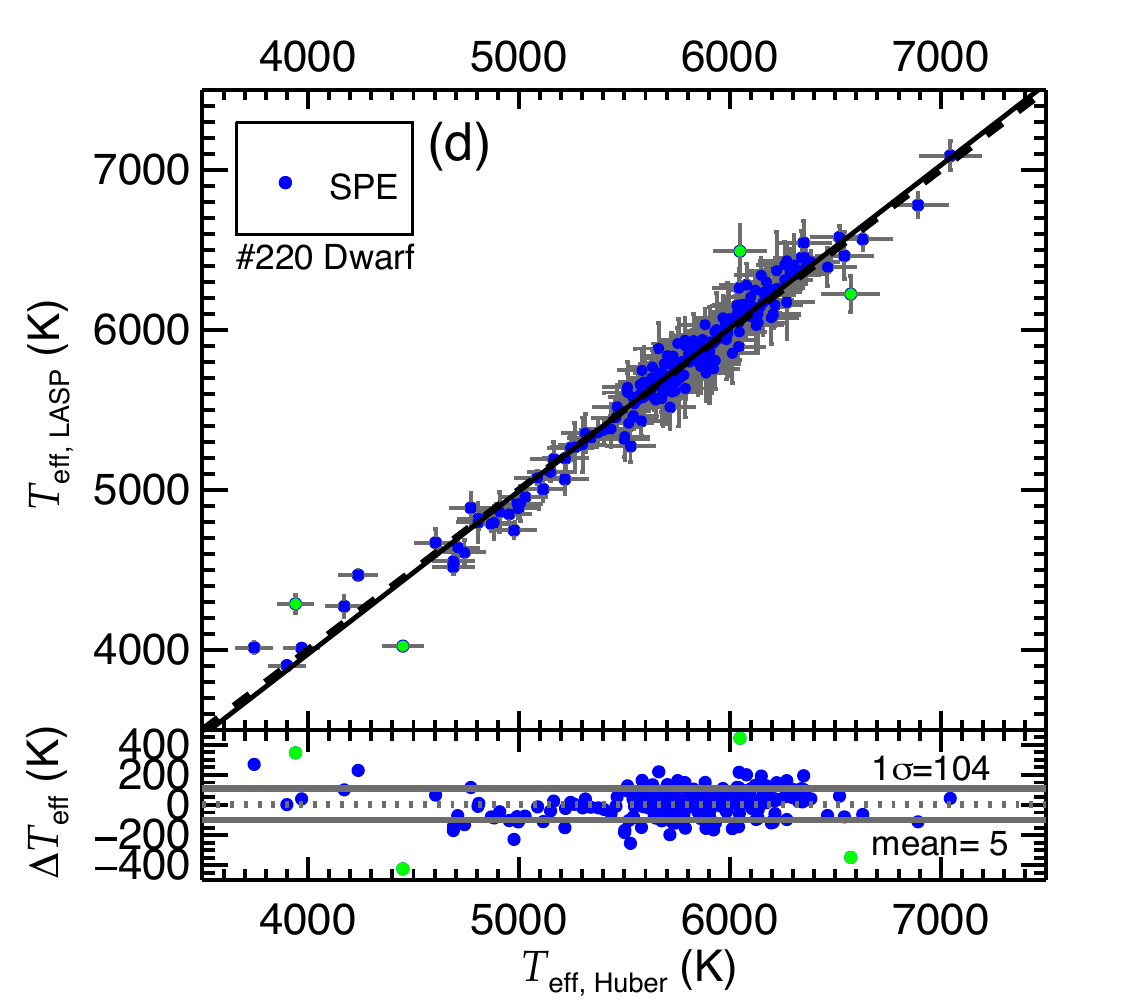}\hfill
   \includegraphics[width=0.333\textwidth]{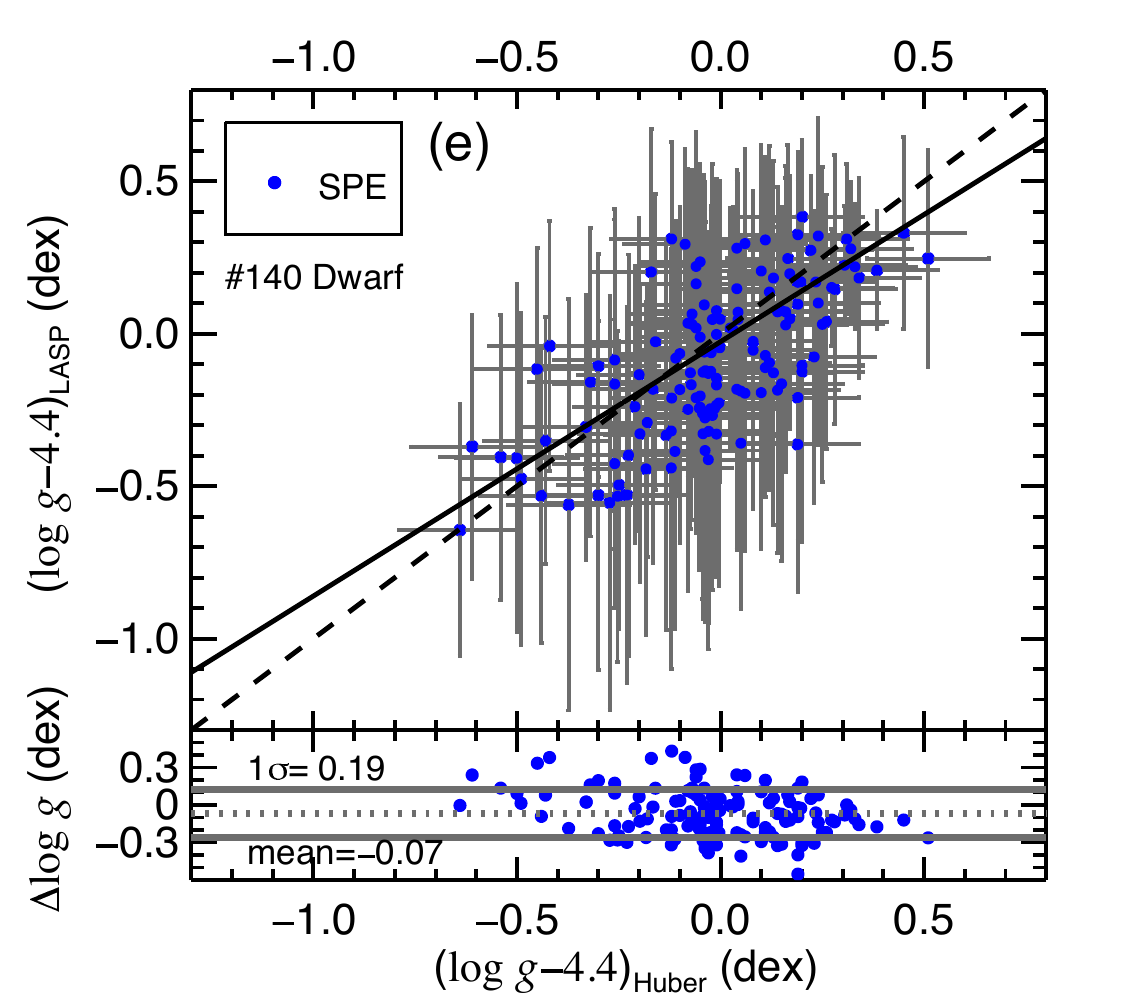}\hfill
   \includegraphics[width=0.333\textwidth]{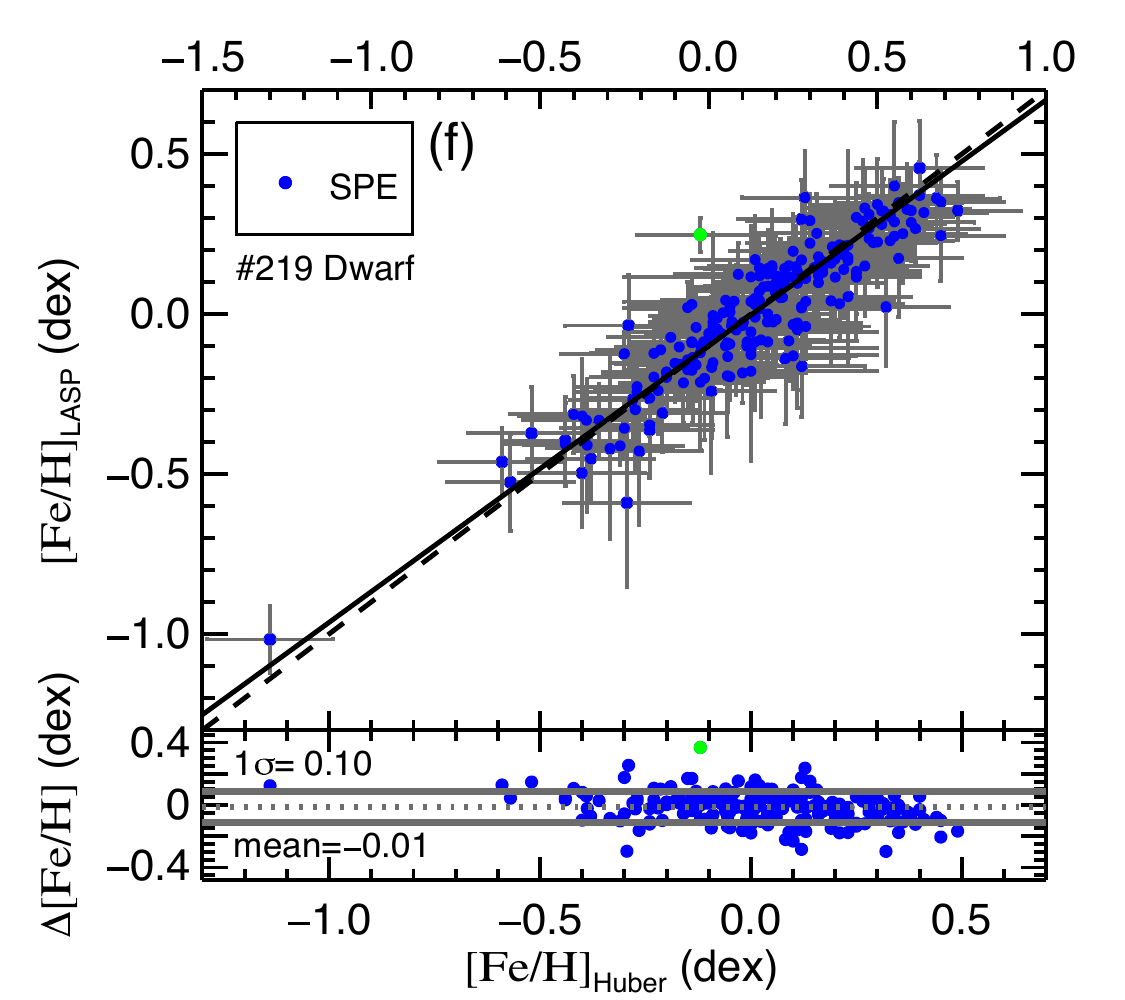}\\
   \includegraphics[width=0.333\textwidth]{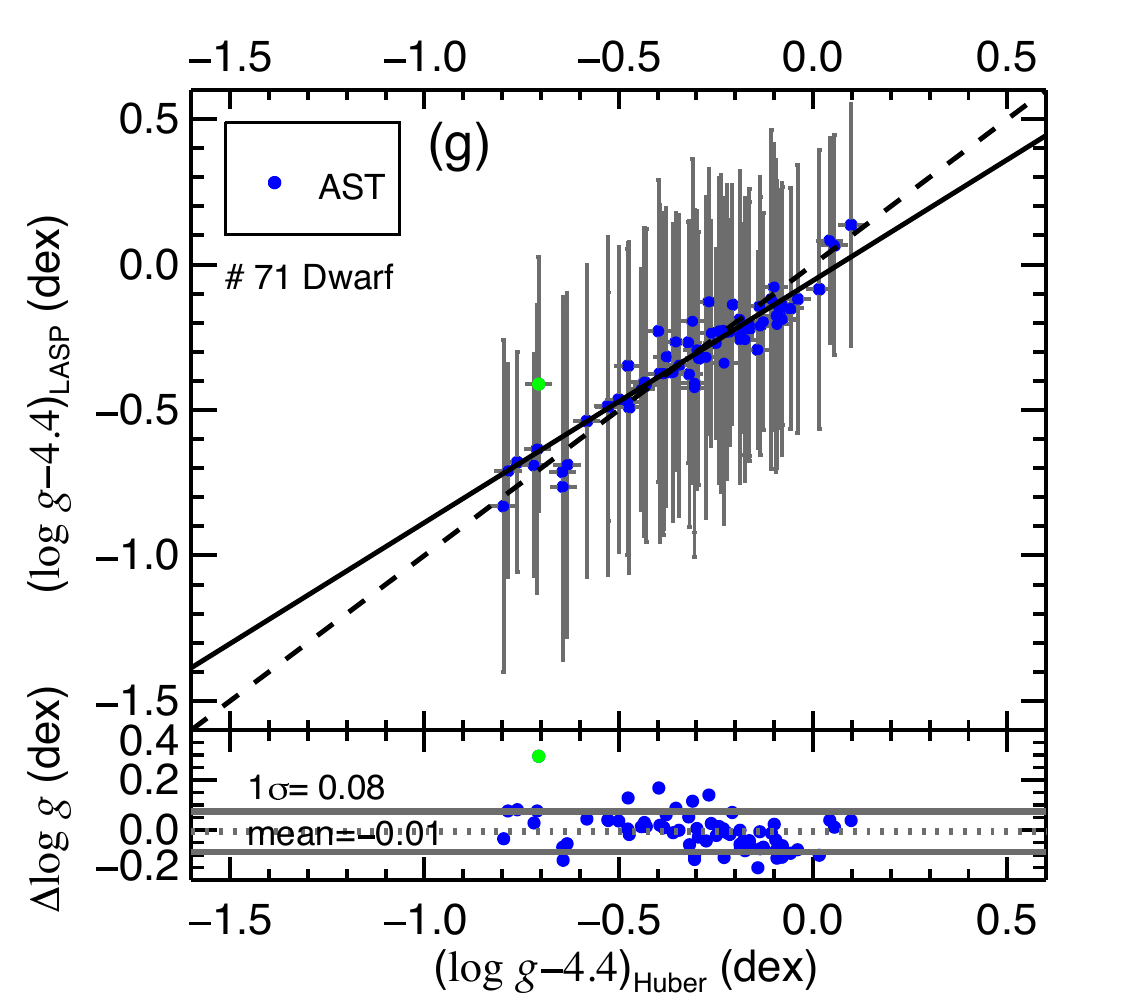}\\
\caption{\footnotesize{Comparison of the \teff, \logg\ and \feh\ as determined by \lasp\ (\lasp, Y-axis of the upper panels) with the subset for which the stellar parameters were derived by the method of spectroscopy (SPE) and asteroseismology (AST) in HT4 (categories C1, C2, C3, C7, C8, and C9 of their Table 1; X-axis). The parameters derived from high resolution spectra are mainly based on 8 works \citep[]{Bat2013,Bru2012,Buc2012,Hub2013,Man2012,Mol2013,Pet2013,Thy2012}. In the upper panels, the stars in common between the two data sets are plotted with their error bars. The black dashed line visualises the bisector while the black solid line is a linear fit to the datapoints. In the lower panels, the difference (\lasp\ - Huber) for these parameters are plotted on Y-axis. The mean bias is indicated by the dotted line and the grey solid line is the 1 $\sigma$ deviation. The results for the giants and dwarfs are given in red and blue, respectively. The stars for which the values are found outside the 3$\sigma$ region around the mean difference are given in green.} \label{fig4}}
\end{figure}

For the external calibration of the \lasp\ results, we used the atmospheric parameters for stars in the \kepler\ field based on $Spectroscopy$ and $Asteroseismology$ as given by H14 (categories C1, C2, C3, C7, C8, and C9 of their Table 1) as the most reliable reference. Given the large dispersion in the methods used for the parameter determination of the literature values originating from \citet{Uyt2011} (indicated with `SPE4' in H14), these values were excluded. The remaining sample of stars was cross-identified with those listed in Table \ref{Tab3} to determining the common stars. Based on the choice of the classification for the dwarfs and giants in H14, these common stars were divided into giant stars (\logg\ $<$ 3.5) and dwarf stars (\logg\ $\geq$ 3.5) to verify the reliability of the \lasp\ parameters of both groups individually. For all the stars in this common subset, we select  the \teff\ and \feh\ values given by H14 that were derived from high-resolution (R $\geqslant$ 20,000) spectroscopic data with analysis pipelines such as SME (Spectroscopy Made Easy; \citealt{Val1996}), SPC (Stellar Parameter Classification; \citealt{Buc2012}), VWA (Versatile Wavelength Analysis; \citealt{Bru2010}), and ROTFIT \citep{Fra2003}. The values of \logg\ were obtained from asteroseismology and spectroscopy. The comparison of atmospheric parameters for the giants and dwarfs in common between the select sub-sample of H14 and the catalogue of the \project\ is presented in Figure \ref{fig4}.

Figure \ref{fig4}(a) shows the comparison of \teff\ for giant stars. The full line in the upper panel gives the result of a linear fit to the data:
\begin{equation}  
T_{\rm eff, LASP}=(0.94 \pm 0.02)T_{\rm eff, Huber}+(299 \pm 104)  \ (\rm for\ gaints) \: ,
\end{equation} 
\noindent
where the subscript `\lasp' indicates the parameters that are derived by \lasp\ from \lamost\ spectra while the subscript `Huber' means that the parameters originate from Table 4 of \citet[hereafter HT4]{Hub2014}. The slope slightly smaller than unit, the best fit indicates that the \teffL\ values are slightly higher than those given by H14 at the low temperature region while it is the other way around at the high temperature region. The values derived from the \lamost\ low-resolution spectra correlate well with those obtained from the high-resolution spectra in the \teff\ range of 3,800~K to 5,300~K. The mean difference and the standard deviation $\sigma$\teff\ of the residuals ($\Delta$\teff\ = \teffL\ - \teffH) amount to 34 K and 131 K, respectively. Only one targets KIC11559263, classified by \lasp\ as a G1 star with the effective temperature 5780 $\pm$ 66 K, is located outside the 3$\sigma$\ region around the mean difference. We checked this \lamost\ spectrum and found that it's quality is very good: the \snrg\ reaches 150 and the error of \teff\ is smaller than the one given in the literature. In general, the errors of \teff\ from the \lasp\ results are smaller than the typical adopted uncertainties as given by H14. The comparison of the \teff\ values for dwarf stars is shown in Figure \ref{fig4}(d). Compared to the result of the giant stars, a tighter relation is found between the \teff\ values for the dwarf stars:
\begin{equation}  
T_{\rm eff, LASP}=(1.02 \pm 0.02)T_{\rm eff, Huber}+(99 \pm 93)  \ (\rm for\ dwarfs) \:
\end{equation}
\noindent
in the \teff\ range of 3,900 K to 7,000 K, although 4 stars (green points) deviate from the 3 times $\sigma$ region around the mean difference. These stars are KIC04832837, KIC12644769, KIC12454461 and KIC08346342 from low to high temperature, respectively. KIC04832837 and KIC12644769 are classified as K5 and K7 type stars with a temperature of 4288 $\pm$ 56 K and 4024 $\pm$ 21 K, respectively. Their spectra have a good quality with  \snrg=15 and 64, respectively.  A problem with the background subtraction results in negative fluxes and a wrong combination of the blue and red channel for the \lamost\ spectrum of KIC12454461. Hence, the \lasp\ \teff\ derived from this spectrum, even if it has a high SNR (\snrg=77), is unreliable. The \lasp\ value of KIC08346342 (a F6 type star with \teff=6224 $\pm$ 111 K) is more trustworthy given the high quality of its \lamost\ spectrum and its error below the uncertainties given by H14. 

The comparison of the \logg\ values that are calculated by the asterseismic method for giant stars is given in Figure \ref{fig4}(b). The linear relation of the \logg\ between the \lasp\ and Huber is described by the following function:
\begin{equation}  
(\log g - 2.5)_{\rm LASP}=(0.80 \pm 0.12)(\log g - 2.5)_{\rm Huber}+(0.17 \pm 0.06)  \ (\rm for\ gaints) \: .
\end{equation}
\noindent
The mean difference and standard deviation $\sigma$\logg\ of the residuals ($\Delta$\logg\ = \loggL\ - \loggH) amount to 0.13~dex and 0.19~dex, respectively. The \loggL\ values tend to be slightly higher than the published values, especially in the range of (\logg$ - 2.5$)$_{\rm Huber}$ $< 0.5$ dex for most giant stars. \citet{Liu2014} estimated the surface gravity of \lamost\ giant stars by using the support vector regression model based on the \kepler\ measured seismic surface gravities given by H14. They also revealed a systematic overestimation of \loggL\ for some giant stars (\logg\ $<$ 4.0 dex and \feh\ $>$ -0.6 dex). The comparisons of the \logg\ derived from spectroscopy (SPE) and asteroseismology (AST) for the dwarfs are shown in Figure \ref{fig4}(e) and (g), respectively. The data are fitted slightly better by a global linear relation than Figure \ref{fig4}(b) (full black line). We obtained the following linear fit as the best approximation for the relation of the \logg\ values for the 211 common dwarf stars: 
\begin{equation}  
(\log g - 4.4)_{\rm LASP}=(0.86 \pm 0.11)(\log g - 4.4)_{\rm Huber}-(0.03 \pm 0.03)  \ (\rm for\ dwarfs) \: .
\end{equation}
\noindent
We still can find the overestimation of \logg\ for the dwarf stars with (\logg$ - 4.4$)$_{\rm Huber}$ $<$ -0.4 dex. Although \lasp\ slightly underestimates the overall \logg\ values for the dwarf stars, there is an overestimation trend towards decreasing \logg\ values. Note that the errors of \lamost\ surface gravities are almost always larger than the uncertainties given by H14 because of the high-precision of \logg\ derived with the method of asteroseismology \citep{Mor2012,Cre2013,Eps2015} and from high-resolution spectra. The mean difference and standard deviation of $\Delta$\logg\ are -0.07 $\pm$ 0.19 dex and -0.01 $\pm$ 0.08 dex for dwarf stars as compared to the values derived from high-resolution spectra and by the method of asteroseismology, respectively. For all dwarf stars, the mean difference and standard deviation of $\Delta$\logg\ is -0.05 $\pm$ 0.16 dex. The F5 star KIC07800289 (\logg=3.99 $\pm$ 0.44 dex) is located out the 3$\sigma$\ region around the mean difference even though its \lamost\ spectrum is of excellent quality (\snrg = 156.9). It is shown as green point in Figure \ref{fig4}(g). 

The linear correlations 
\begin{equation}  
\rm [Fe/H]_{\rm LASP}=(0.95 \pm 0.06)[Fe/H]_{\rm Huber}-(0.05 \pm 0.03)  \ (\rm for\ gaints) \: and
\end{equation}
\begin{equation}  
\rm [Fe/H]_{\rm LASP}=(0.96 \pm 0.06)[Fe/H]_{\rm Huber}-(0.00 \pm 0.01)  \ (\rm for\ dwarfs) \: ,
\end{equation}
\noindent
which are close to the 1:1 relation, are found as the best fit for the relation between the [Fe/H] values for the common giant and dwarf stars and are plotted  are plotted as a black full line in Figure \ref{fig4}(c) and (f), respectively. The majority of the residual \feh\ values ($\Delta$$\feh\ = \fehL\ - \fehH$) are concentrated around 0~dex. Only a small fraction of them are found outside the 3$\sigma$ $\Delta$\feh\ region.  The reliability of the \feh\ derived from the \lamost\ low-resolution spectra with the \lasp\ have already received recognition by comparison with results derived from the high-resolution spectra in a relatively small metallicity range of -0.3 dex to +0.4 dex \citep{Don2014}. Our comparison shows that the \feh\ values obtained with \lasp\ are reliable in the range between -1.0~dex and 0.5~dex for giant stars, and between -0.6 dex and 0.5 dex for dwarf stars. We note that \lasp\ underestimates \feh\ for the giant stars in the range \feh\ $>$ 0.2 dex. The mean differences and standard deviations of $\Delta$\feh\ are -0.04 $\pm$ 0.15 dex and -0.01 $\pm$ 0.10 dex for giant and dwarf stars, respectively. The discrepant object, the value of $\Delta$\feh\ deviate from the 3$\sigma$ region around the mean difference, shown as a green point in Figure \ref{fig4}(f) is KIC10318874 (\feh = 0.25 $\pm$ 0.05 dex, \snrg = 85). It has been classified by \lasp\ as a F5 dwarf star. The \lamost\ spectrum of this star is of good quality and the error on the metallicity is relatively low, which are both in favour to consider the \lasp\ results as trustworthy. 

Comparisons between the \lasp\ parameters and those listed in the considered sub-sample of HT4 illustrate that the correlation of these parameters is close to the 1:1 relation. This result favors the reliability of the \lasp\ determinations in wide ranges: from 3,800~K to 6,600~K for \teff, from 1.5~dex and 4.9~dex for \logg\ and from -1.0~dex to 0.5~dex for \feh. Note that the reliable ranges of atmosphere parameters are not completely the same for giant and dwarf stars. We need more stars with values outside those ranges to validate the reliability of the \lasp\ results. The atmospheric parameters of the dwarfs in the \project\ are superior to the values of the giants in general as shown in the Figure \ref{fig4}. 

\subsection{Internal Calibration}
\label{sect:5.2}
In total, 7 LK-fields have been observed more than once from 2012 to 2014 as given in the upper part of Table \ref{Tab1}. Moreover, there is overlap between adjacent LK-fields to allow a full coverage of the \kepler\ field as shown in Figure \ref{Fig1}. Hence, we obtained more than one spectrum for a substantial fraction of stars: 7,550 stars have been observed two times, 986 stars three times, 92 stars four times, and 5 stars at least five times (bottom part of Table \ref{Tab1}). The atmospheric parameters of these stars have been derived from the spectra of multiple observations. 

\begin{figure} 
\epsscale{0.6}
\plotone{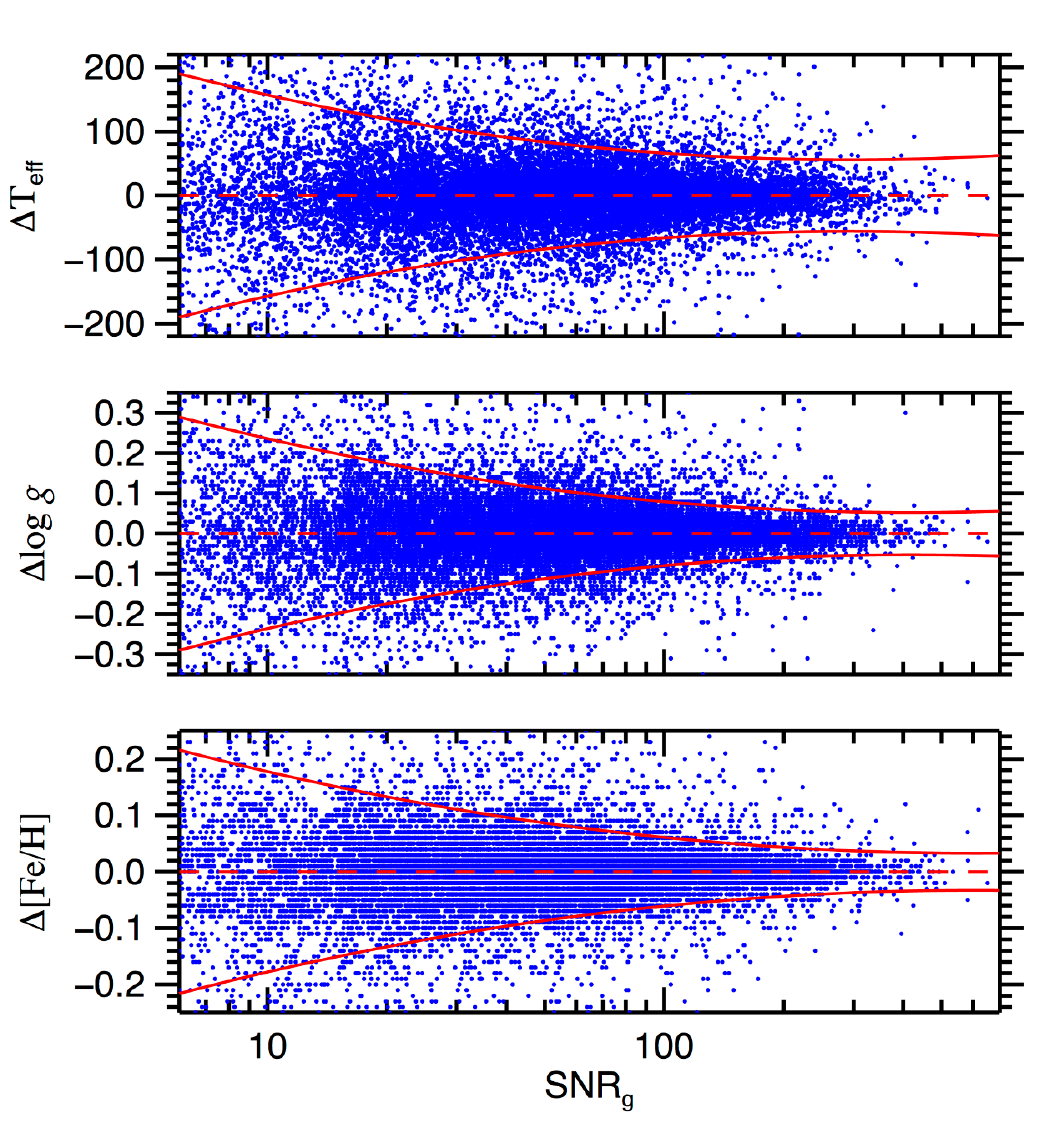}
\caption{The unbiased estimation for the multiple observation targets as a function of \snrg\ (blue dots). The 1$\sigma$ confidence levels are fitted with a second-order polynomial (red solid lines).  \label{fig5}}
\end{figure}

We assess the internal errors of stellar atmospheric parameters by making comparisons of the atmospheric parameters obtained from different \lamost\ spectra of the same objects. We used the method of the unbiased estimator:
\begin{equation}  
 \Delta P_{i}=\sqrt{n/(n-1)}(P_{i}-\bar{P}) \: ,
 \end{equation} 
\noindent 
with $ i $ = 1, 2, ..., $ n $, where $ i $ is one of the individual measurements and $ n $ is the total number of measurements for parameter $ P $ \citep{Xie2016}. The values of the unbiased estimations for the effective temperature ($\Delta$\teff), surface gravity ($\Delta$\logg) and metallicity ($\Delta$\feh) versus \snrg\ of the \lamost\ spectra are shown in Figure \ref{fig5}, respectively.
Second-order polynomials as a function of \snrg\ are used to fit the 1$\sigma$ confidence interval in various bins of \snrg. The following fitting functions are given as red solid lines: 
\begin{eqnarray} 
\left\{ \begin{array}{l}
  \sigma_{T_{\rm eff}}=47.0X^{2}-232.1X+342.1\ $K$ \\
  \sigma_{\log g}=0.070X^{2}-0.366X+0.532\ $dex$ \\
  \sigma_{\rm [Fe/H]}=0.045X^{2}-0.253X+0.386\ $dex$ \\
\end{array} \right.
\end{eqnarray}

\noindent
where $X$ is the base-10 logarithm of \snrg. The internal errors of the parameters with the \snrg\ $\geq$ 6.0 are 91 K, 0.12 dex and 0.09 dex for \teff, \logg\ and \feh, respectively.  For the stars with \snrg\ $\geq$ 50, the inner errors of them are 68 K, 0.08 dex and 0.06 dex, respectively. 

\subsection{Calibration of \lasp\ Stellar Parameters}
\label{sect:5.3}
\begin{figure} 
\centering
   \includegraphics[width=0.5\textwidth]{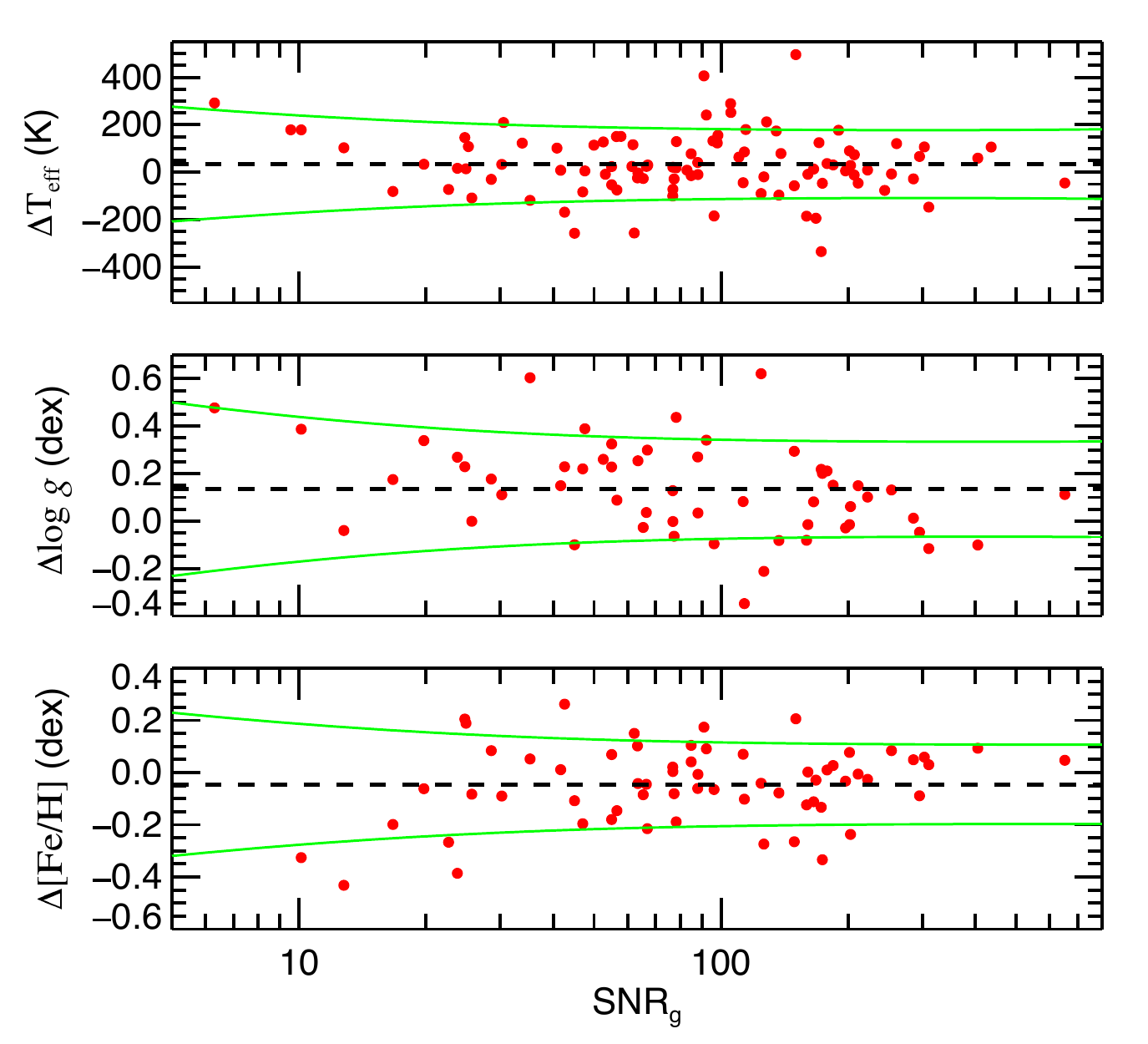}\hfill
   \includegraphics[width=0.5\textwidth]{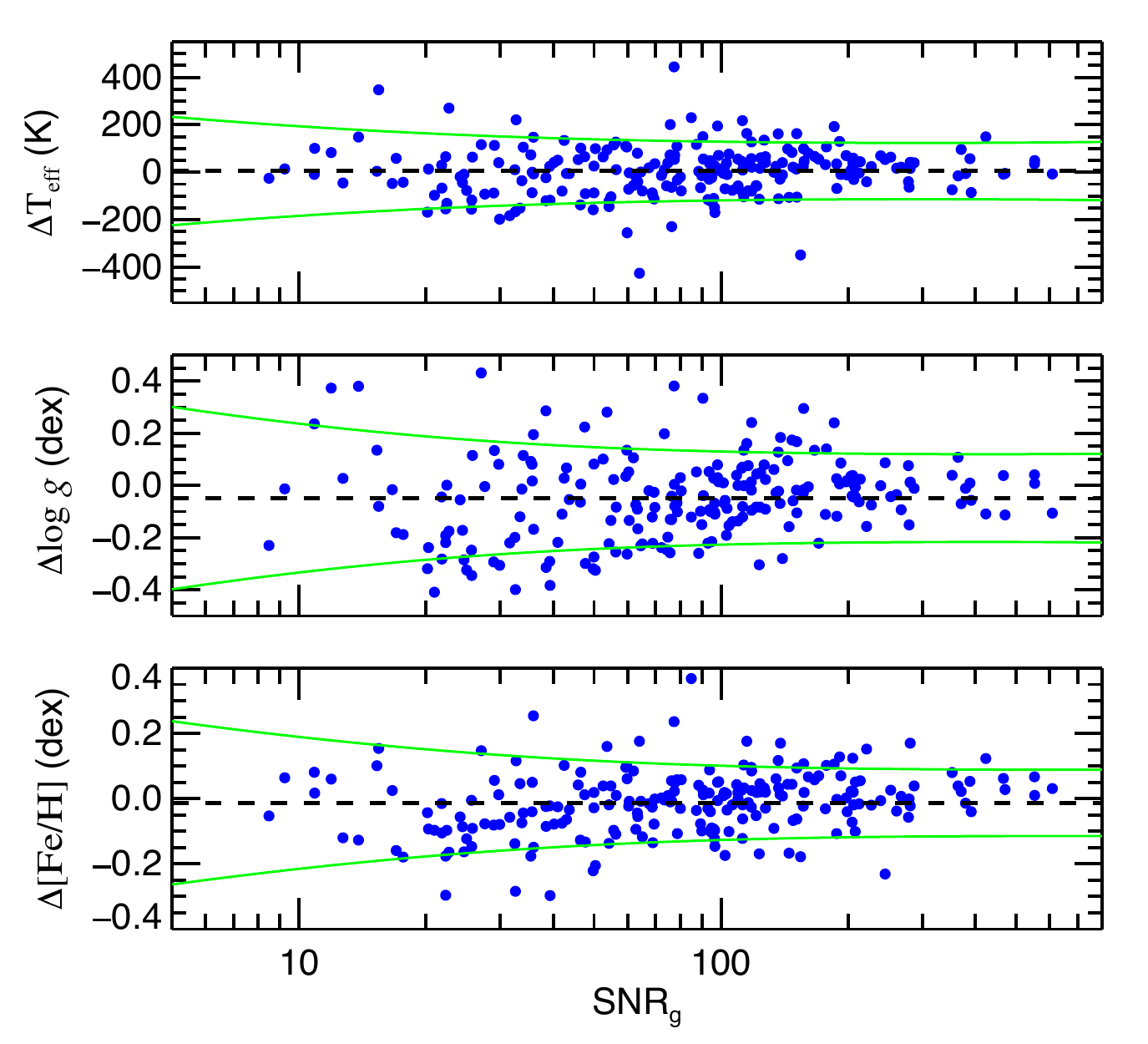}\\
\caption{The dispersions of the \lasp\ stellar atmospheric parameter uncertainties (the green lines), taking both intern and external uncertainties into account, are calculated for the giants (left panels) and dwarfs (right panels), respectively. The mean biases are indicated by the black dashed lines when compared with the published values in HT4, the red points and blue points represent the relation between the differences (\lasp\ - Huber) and \snrg\ for the common giants and dwarfs, respectively. \label{fig6}}
\end{figure}

The external calibration of the \lasp\ stellar parameter uncertainties for giants and dwarfs are obtained by a separate comparison with the published values in HT4 as described in Section \ref{sect:5.1}. The internal errors of the stellar atmospheric parameters as a function of the base-10 logarithm of \snrg\ are calculated using the parameters of multiple observations objects as described in Section \ref{sect:5.2}.  The dispersions of the \lasp\ stellar atmospheric parameter uncertainties (the green lines), taking both intern and external uncertainties into account, are calculated for the giants (left panels) and dwarfs (right panels) in Figure \ref{fig6}, respectively. The relation between the differences (\lasp\ - Huber) and \snrg\ are shown for the common giants (red points) and dwarfs (blue points) when compared with the published values in HT4 as described in Figure \ref{fig4}. The black dashed lines describe the mean biases of the parameters. 

The external and internal uncertainties of the \lasp\ parameters are combined to redefine the systematic deviation and the errors of stellar atmospheric parameters within the \project. The calibrated relations are given by:
\begin{eqnarray} 
\left\{ \begin{array}{l}
         P_{i}=(P_{i, \rm LASP} - a)/b \\
         \sigma=\sqrt{\sigma_{in}^{2}+\sigma_{ex}^{2} } \\
\end{array} \right.
\end{eqnarray}

\noindent
where for each observation $i$, $P_{i}$  denotes the calibrated stellar parameters and $P_{i, \rm LASP}$ is the \lasp\ value while $a$ and $b$ are the zero and slope of the linear functions as given in Section \ref{sect:5.1}. $\sigma$ is the calibrated errors of atmospheric parameters. $\sigma_{in}$ is the inner deviation described by a second-order polynomial as a function of \snrg\ by fitting the data points of the unbiased estimation for targets with multiple observations (see Section \ref{sect:5.2} \& Figure \ref{fig5}). $\sigma_{ex}$ is the external deviation of the \lasp\ atmospheric parameter uncertainties (see Section \ref{sect:5.1}). Based on these calibrated relations, we recalculated the \lasp\ stellar parameters according to the following formulae for giant and dwarf stars, respectively: 
\begin{eqnarray} 
T_{\rm eff} \left\{ \begin{array}{l}
    \sigma_{in}=47.0X^{2}-232.1X+342.1\ $K$                            \\
    a=299\ $K$, b=0.94;  \sigma_{ex}=131\ $K$ \ ($for giants$)   \\
    a=99\ $K$, \ b=1.02;  \sigma_{ex}=104\ $K$ \ ($for dwarfs$), \\        
\end{array} \right.
\end{eqnarray}
\begin{eqnarray} 
\log g \left\{ \begin{array}{l}
    \sigma_{in}=0.070X^{2}-0.366X+0.532\ $dex$                               \\
    a=0.67\ $dex$, b=0.80;  \sigma_{ex}=0.19\ $dex$\ ($for giants$)    \\ 
    a=0.59\ $dex$, b=0.86;  \sigma_{ex}=0.16\ $dex$\ ($for dwarfs$),  \\
\end{array} \right.
\end{eqnarray}
\begin{eqnarray} 
\rm [FeH] \left\{ \begin{array}{l}
    \sigma_{in}=0.045X^{2}-0.253X+0.386\ $dex$                             \\
    a=0.05\ $dex$, b=0.95;  \sigma_{ex}=0.15\ $dex$\ ($for giants$)  \\ 
    a=0.00\ $dex$, b=0.96;  \sigma_{ex}=0.10\ $dex$\ ($for dwarfs$),\\ 
\end{array} \right.
\end{eqnarray}

\noindent
where $X$ is the base-10 logarithm of \snrg. 
The calibrated atmospheric parameters and their errors are listed in Table \ref{Tab4}, which includes `Obsid', `Target' and the information of the spectra as listed in Table \ref{Tab3}. 
\begin{deluxetable}{rrrllrr} 
\tabletypesize{\scriptsize}
\tablecaption{The catalog of calibrated \lasp\ stellar atmosphere parameters in the first round of observations for the \project.  \label{Tab4}}
\tablewidth{0pt}
\tablehead{
\colhead{Obsid} & \colhead{Target} & \colhead{\snrg}  & \colhead{Subclass}  & \colhead{\teff\ (K)}  & \colhead{\logg\ (dex)} & \colhead{\feh\ (dex)}\\} 
\startdata                                                                                                     
  52201011  & KIC07042868 &    76.73  &     G5 &  4844$\pm$149 &   2.485$\pm$0.211 &   0.031$\pm$0.165 \\
  52201018  & KIC06957157 &    50.72  &     G5 &  4727$\pm$155 &   2.284$\pm$0.220 &   0.051$\pm$0.173 \\
  52201025  & KIC06957977 &    47.57  &     K1 &  4758$\pm$156 &   1.994$\pm$0.222 &  -0.189$\pm$0.174 \\
  52201040  & KIC07206837 &    46.50  &     F6  &  6183$\pm$134 &   4.167$\pm$0.198 &   0.230$\pm$0.134 \\
  52201060  & KIC07368371 &    98.09  &     K3 &  4114$\pm$146 &   1.705$\pm$0.206 &  -0.121$\pm$0.162 \\
  \ldots         &  \ldots              &     \ldots & \ldots &   \ldots                &   \ldots                    &   \ldots                    \\
 250016240 & KIC12933001 &    25.87  &     K1 &  4802$\pm$149 &   3.967$\pm$0.223 &   0.044$\pm$0.155 \\
 250016244 & KIC12983407 &    20.10  &     G8 &  5448$\pm$158 &   4.386$\pm$0.236 &  -0.019$\pm$0.166 \\
 250016245 & KIC12883530 &  102.95  &     F0  &  7022$\pm$122 &   4.003$\pm$0.178 &  -0.003$\pm$0.116 \\
 250016248 & KIC12933571 &    19.47  &     K5 &  4448$\pm$178 &   2.510$\pm$0.259 &  -0.227$\pm$0.202 \\
 250016249 & KIC12883443 &    18.22  &     K0 &  5086$\pm$161 &   3.742$\pm$0.242 &   0.036$\pm$0.171 \\
\enddata
\end{deluxetable}

\section{Statistical Analysis of Stellar Parameters}
\label{sect:6}
For 8,633 targets, more than one spectrum of \lamost\ observation have been analyzed (see bottom rows of Table\,\ref{Tab1}). We determined the stellar parameters from the multiple \lamost\ spectra with different SNR values. The accuracy and credibility of these parameters mainly depends on the quality of the observed spectra. For these stars with multiple observations, we refer to the parameters that are derived from the \lamost\ spectrum with the highest \snrg\ in what follows (unless stated otherwise). We provide a general statistical analysis of stellar parameters for all 51,399 \kepler\ stars in the \project.

\begin{figure} 
\epsscale{0.80}
\plotone{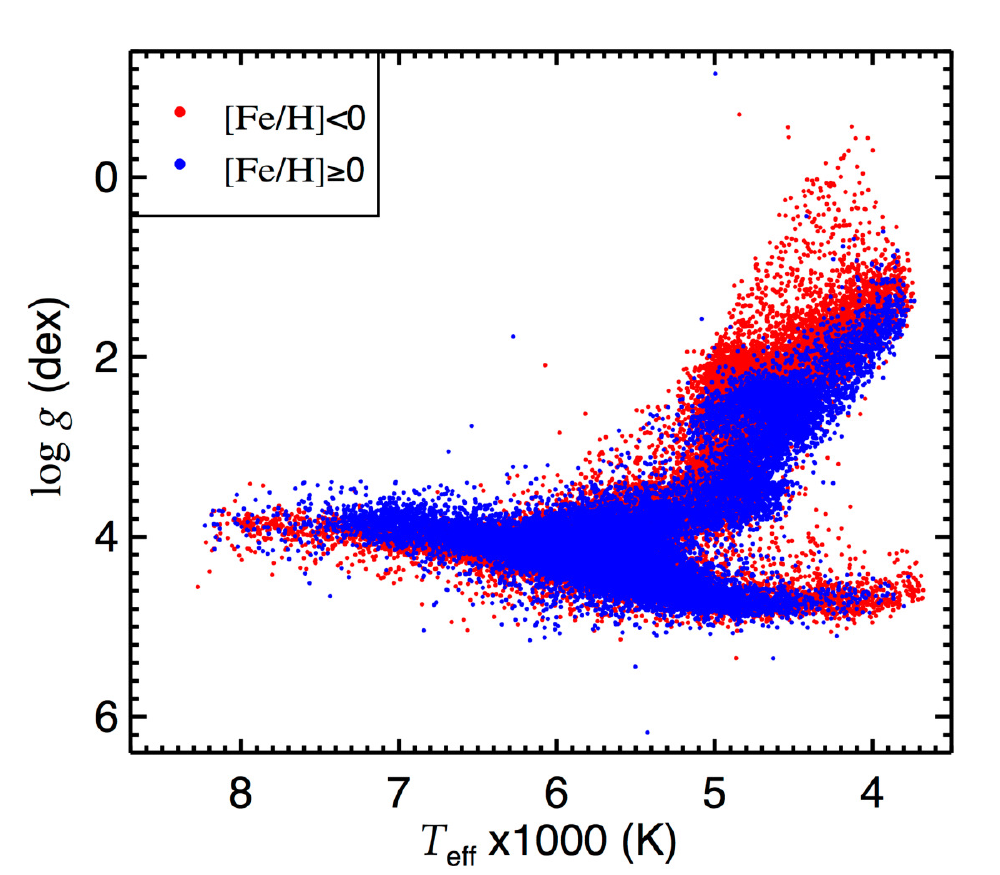}
\caption{The position of the analyzed targets in a Kiel diagram (\logg\ vs. \teff) based on the calibrated stellar atmospheric parameters derived with \lasp\ from the spectra obtained in the \project.  The colors denote different values of \feh\ (see legend in the top left corner). \label{fig7}}
\end{figure}

The calibrated \lasp\ stellar atmospheric parameters (\teff, \logg\ and \feh) and the \lasp\ \vrad\ are found in the ranges 3,678 $\sim$ 8,275 K, -1.150 $\sim$ 6.174 dex,  -2.811 $\sim$ 1.105 dex and -472 $\sim$ 120 km/s, respectively. The mean errors of the measured stellar parameters are 2.75\% in \teff, 0.215 dex in \logg, 0.152 dex in \feh, and 18~\kms\ in \vrad. In Figure \ref{fig7}, the analyzed stars are plotted in the \teff\ - \logg\ diagram (so-called `Kiel' diagram). As can be seen, they are mainly located in the main sequence and the classical instability strip. 

\begin{figure} 
\epsscale{0.80}
\plotone{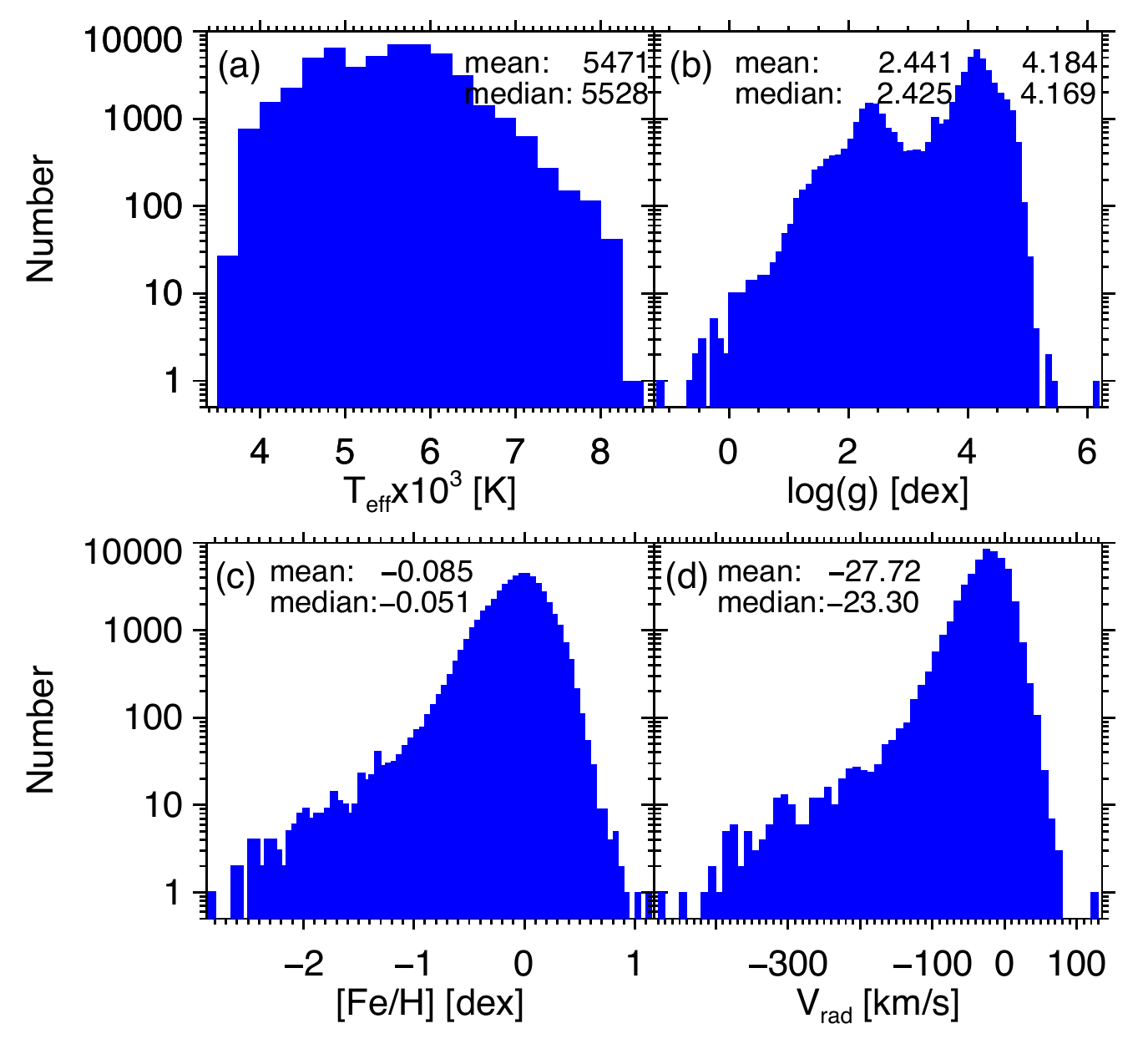}
\caption{The histogram distribution of the 51,399 \kepler\ targets in (a) \teff\ (250 K per bin), (b) \logg\ (0.1 dex per bin), (c) \feh\ (0.05 per bin) and (d) \vrad\ (10 km/s per bin). We calculate a set of mean and median values of these parameters. Two sets of mean and median values of the \logg\ are given for the two parts with a cut-off point at \logg\ = 3.5 dex. \label{fig8}}
\end{figure}

Figure \ref{fig8} shows the histograms of \teff, \logg, \feh\ and \vrad, which helps to easily find the stars of particular interest in the \project, such as metal-poor stars \citep{Li2015apj,Li2015pa,Aok2014,Wu2010} and high velocity stars \citep{Zhong2014}. Two distinct peaks, which is a reflection of the main-sequence and the red giant branch, are shown in the histograms of \teff\ and \logg\ in Fig. \ref{fig8}(a) and (b), respectively. The distribution of \feh, as shown in Fig. \ref{fig8}(c), peaks near solar metallicity but is left-skewed with a tail that stretches away from the centre. As metal-deficient stars provide fundamental information of the chemical abundance formation and evolution from the early stage of the galaxy \citep{Li2010,Den2012}, they are regarded as fossils of the early generation of stars. Based on the nomenclature for stars of different metallicity \citep{Bee2005} and the visual scan through 475 spectra with \feh\ $<$ -1.0 dex, we finally classified 106 targets as candidate metal-poor stars (MPs, \feh\ $<$ -1.0 dex) and 9 targets as candidate very metal-poor stars (VMPs, \feh\ $<$ -2.0 dex). As shown in Fig. \ref{fig8}(d), the peak of the \vrad\ distribution deviates from 0~\kms and the mean value of the \vrad\ is -27.72~\kms. After scanning visually the spectra of 62 stars with \vrad\ $<$ -300 km/s in the catalogue of the \project, only 18 stars were classified as candidate high-velocity stars (HVs) to study their natures and formation mechanisms, and to help improving our understanding of the structural properties of the galaxy. All the that are classified as a particular object are listed in the Table \ref{Tab5}.

Note that most targets whose parameter values meet the conditions of the particular objects (MPs, VMPs and HVs) were rejected as candidates after a visual inspection of the data mainly due to the problems of the \lamost\ spectra as already mentioned in \citet{Gra2016}. For most \lamost\ spectra, the background subtraction could be improved, but some spectra still exhibit negative flux values in some of their spectral lines. For spectra with an insufficient background subtraction, the determination of the parameters will also be inaccurate. Some spectra have low SNR values due to the low observed fluxes which makes it difficult to use the noisy spectral features for stellar parameter determinations. All the stars with \lamost\ spectra suffering from at least one of these issues could not be retained as a candidate peculiar object. Most extreme values are usually derived from the problematic spectra which are ascribed to the sorts of problems as given above. For this reason, the relative number of `wrong' values for the parameters would be higher if we are looking at the extreme values (like the extremely low metallicities and extremely fast moving stars). Just because such stars are rare and we don't expect many of them, and hence such values in the results can be indicative of errors in the analysis. The latest \lamost\ reduction and analysis pipeline has been applied in an attempt to solve the problems. Satisfyingly, we estimate that only about 2\% spectra are affected by the above mentioned problems in the \project\ after the quality of the spectra in four selected plates, which are observed on 2013/05/22, 2013/10/17, 2014/05/20 and 2014/09/17, have been visually checked.

\section{Comparison with KIC}
\label{sect:7}
The KIC parameters were mostly derived from the multi-band photometric observations using a set of Sloan filters. This method is less reliable compared to parameter determination from spectroscopic data. 
The shortcomings of the KIC parameters have already been indicated in several previous works \citep{Hek2011,Bro2011,Don2014,Liu2014}. Due to several known systematic defects \citep{Bro2011}, the accuracy of the atmospheric parameters as given in the KIC cannot reach the requirements for asteroseismic studies in many cases. As the main goal of the \project\ is to provide more accurate stellar parameters for objects in the KIC, we examined the reliability of stellar atmospheric parameters in the KIC by comparing them to the calibrated parameters that we obtained for stars in the \project. 

\begin{figure} 
\centering
       \includegraphics[width=0.333\textwidth]{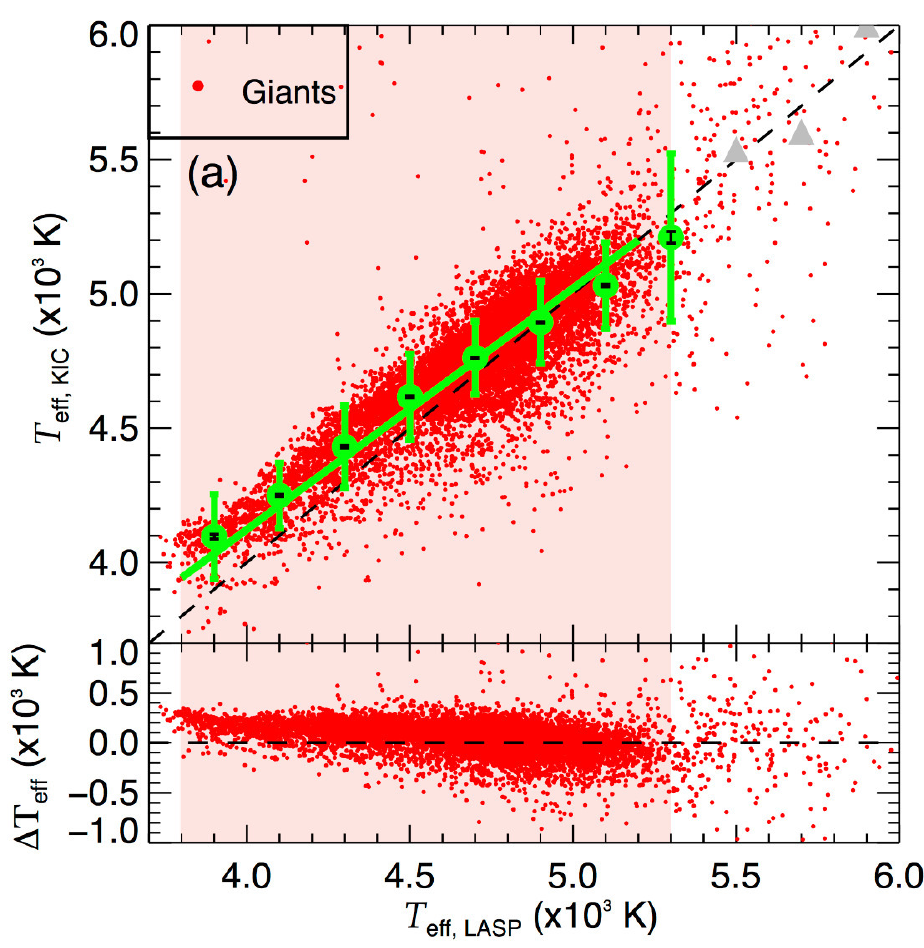}\hfill
       \includegraphics[width=0.333\textwidth]{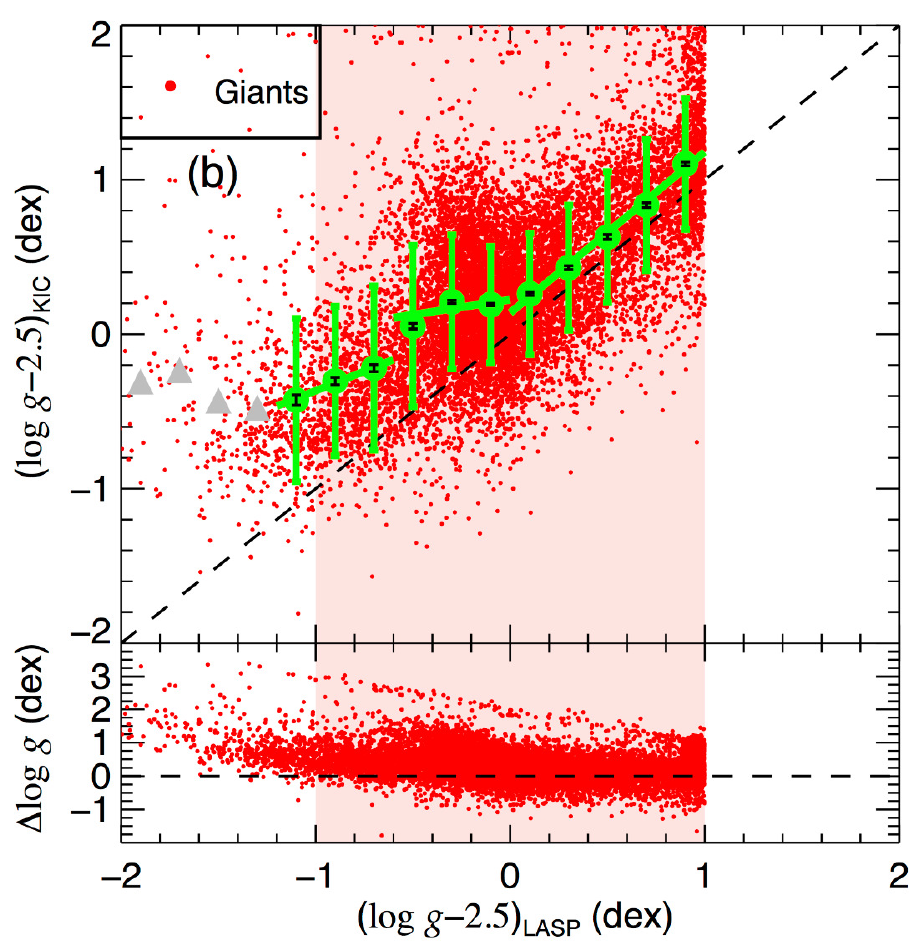}\hfill
       \includegraphics[width=0.333\textwidth]{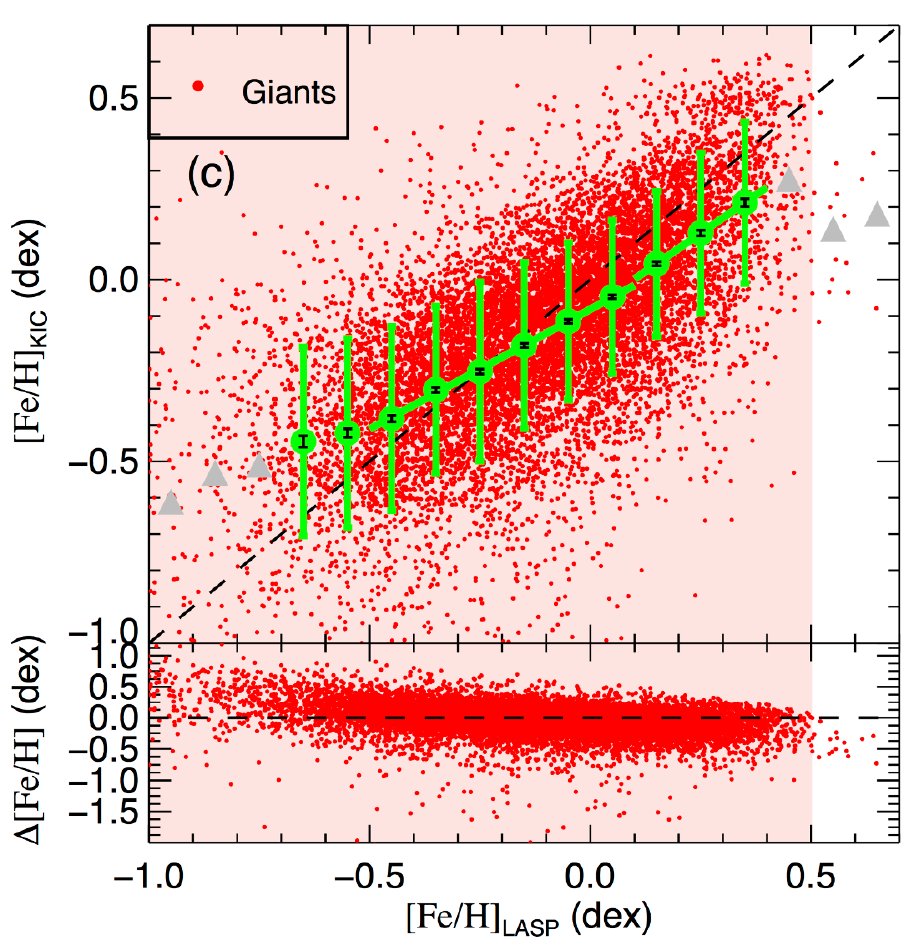}\\
       \includegraphics[width=0.333\textwidth]{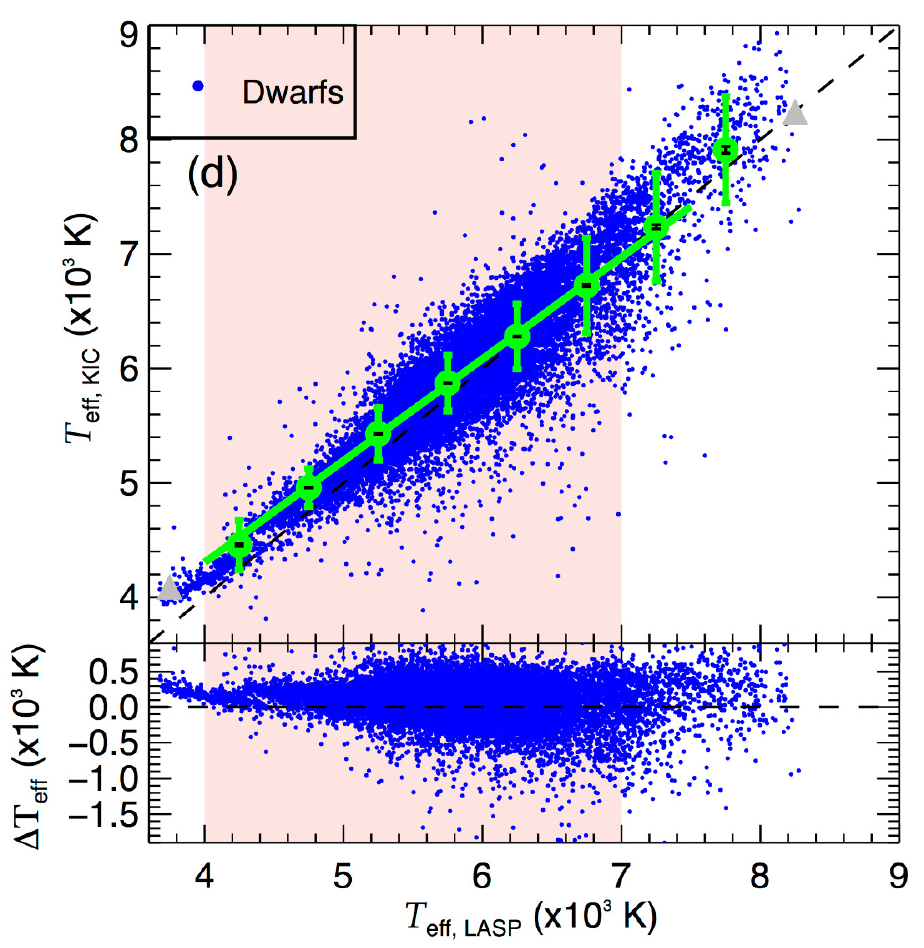}\hfill
       \includegraphics[width=0.333\textwidth]{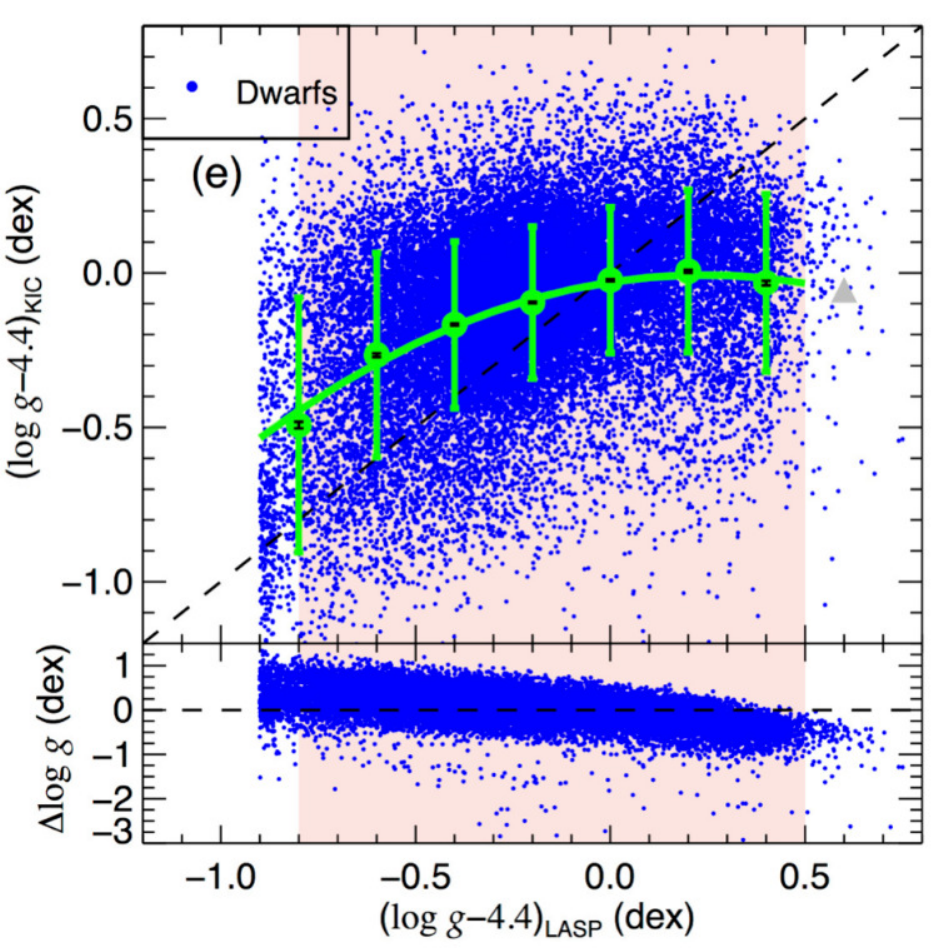}\hfill
       \includegraphics[width=0.333\textwidth]{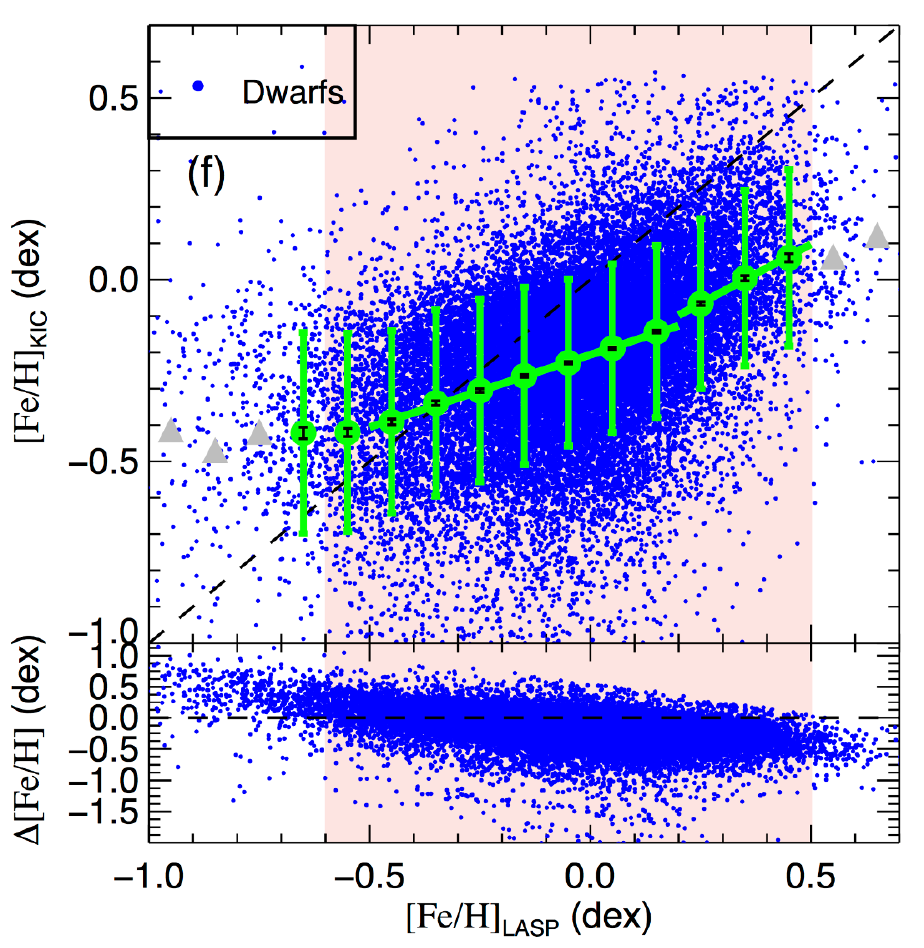}\\
 \caption{Comparison of the calibrated \teff, \logg\ and \feh\ as determined by \lasp\ (\lasp\, X-axis) with those from KIC (KIC, Y-axis of the upper panels). In the upper panels, the common stars between the two datasets are plotted with red points for giant stars (\logg\ $<$ 3.5 dex) and blue points for dwarf stars (\logg\ $\geq$ 3.5). The green filled circles represent the mean values of each bin with at least 200 data points. The size of each bin is 500 K (for dwarfs) or 200 K (for giants) for \teff, 0.2 dex for \logg\ and 0.1 dex for \feh. The mean values of bins with less than 200 data points are given with grey triangles. The black dashed lines show the 1:1 relations while the green solid line are linear fits to the selected green filled circles. The outer error bars give the standard deviations and the black inner error bars give the standard deviation of the mean. In the lower panels, the difference (KIC - \lasp) for these parameters are plotted in the Y-axis. The zero value is indicated by the black dashed line. The regions for which the reliability of the \lasp\ parameters has been verified by the external calibration in Section \ref{sect:5.1} are indicated with a background in pink. \label{fig9} 
}.
\end{figure}

There are 51,399 targets in the catalog of calibrated \lasp\ stellar atmospheric parameters (Table \ref{Tab4}) that can be identified with an object from the KIC. However, stellar atmospheric parameters are listed in the KIC for only 41,775 of them. 
Figure \ref{fig9} shows the comparisons between the \lasp\ and KIC parameters. It is clear from each of the upper panels that a lot of data points deviate from the 1:1 relation with a large scatter.
 All data points are divided into several bins with the bin size 500 K (for dwarfs) or 200 K (for giants) for \teff, 0.2 dex for \logg\ and 0.1 dex for \feh. If the number of data points in the bin is more than 200, the mean values and standard deviations of these data are represented by filled green circles and outer error bars, respectively. For the other bins, the mean values of the data are given by grey triangles. The black inner error bars give the standard deviation of the mean. The reliable filled green circles were fitted by straight lines or curve in Figure \ref{fig9}.

The linear relation of \teff\ between the \project\ and KIC is satisfactory for most stars, although there are large discrepancies for a small portion of data in Figure \ref{fig9} (a, d). There are no obvious differences in the quality of the \teff\ measurements for dwarf and giant stars except that the KIC overestimates \teff\ for most giant stars in the low temperature range. Even though the reliability of the \lasp\ parameters has not been proven for stars with \teff\ $>$ 7000 K for dwarfs, we still observe a close relation between the \teff\ values of both catalogues. A linear fit of the green points restricted, almost entirely, to the range with reliable \lasp\ \teff\ values (see Section \ref{sect:5.1}; given in pink on Figure \ref{fig9}) leads to the following relations:
\begin{eqnarray} 
\left\{ \begin{array}{l}
T_{\rm eff, KIC} = (0.898 \pm 0.003)T_{\rm eff, LASP} + (0.529 \pm 0.016) \ (\rm for\ gaints) \\
T_{\rm eff, KIC} = (0.889 \pm 0.003)T_{\rm eff, LASP} + (0.748 \pm 0.015) \ (\rm for\ dwarfs) ,
\end{array} \right.
\end{eqnarray}

\noindent
where the subscript `KIC' indicates the values originating from the KIC. 

In Figure \ref{fig9} (b, e), the comparisons of the \logg\ values are given for giant and dwarf stars, respectively. There is a large dispersion around the bisector in the upper panels. Hence, for giants, we divided the filled green circles into three regions to fit these points with a linear relationship, the fitting expression is given as follow:
\begin{eqnarray} 
\left\{ \begin{array}{l}
(\log g-2.5)_{\rm KIC} = (0.502 \pm 0.097)(\log g-2.5)_{\rm LASP} + (0.137 \pm 0.084)\ (\rm \log g\ \epsilon\ [1.3, 1.9]) \\
(\log g-2.5)_{\rm KIC} = (0.193 \pm 0.042)(\log g-2.5)_{\rm LASP} + (0.223 \pm 0.010)\ (\rm \log g\ \epsilon\ [1.9, 2.5]) \\
(\log g-2.5)_{\rm KIC} = (1.042 \pm 0.017)(\log g-2.5)_{\rm LASP} + (0.138 \pm 0.010)\ (\rm \log g\ \epsilon\ [2.5, 3.5]) .
\end{array} \right.
\end{eqnarray}

\noindent
For dwarfs, we fitted all points with a quadratic polynomial in the 3.5-4.9 dex range as follows:
\begin{equation} 
(\log g-4.4)_{\rm KIC} = (-0.406 \pm 0.017)(\log g-4.4)^2_{\rm LASP} + (0.196 \pm 0.008)(\log g-4.4)_{\rm LASP} -(0.030 \pm 0.002) \: .
\end{equation}

\noindent
The scatter of the KIC \logg\ in the individual bins is about 0.426 and 0.292 dex for giants and dwarfs, respectively. The biases of the \logg\ values mainly come from the large uncertainties of \logg\  in KIC. This is mainly due to the shortcomings of the KIC stellar parameters. The primary goal of the KIC is to be able to distinguish the cool giant stars from the dwarf stars but the \logg\ values are not accurate enough for asteroseismic studies \citep{Bro2011}. The KIC \logg\ values are generally larger than the \lasp\ values for most dwarf and giant stars, but KIC underestimate the \logg\ in the range \logg\ $>$ 4.4 dex for most dwarfs, and we can find an obvious deviation belt from 0.5 dex to about 2.5 dex for giant stars in the lower panel of Figure \ref{fig9}(b). 

The situation when comparing \feh\ values between the two datasets is even worse as can be seen on Figure \ref{fig9} (c, f). The scatter of the KIC \feh\ is about 0.223 and 0.242 dex for giants and dwarfs in the individual bins, respectively. This situation should be mainly ascribed to problems of the KIC \citep{Pla2014}. There is also an obvious wide deviation for most giant and dwarf stars: the KIC seriously overestimates the \feh\ values for most metal poor stars, but it underestimates the \feh\ values for most stars around 0 dex, as mentioned in \citet{Don2014}, especially for dwarfs. Finally, we divided the green points into two ranges to separately fit them with a linear function. We give the resulting relations for giants: 
\begin{eqnarray} 
 \left\{ \begin{array}{l}
\rm [Fe/H]_{\rm KIC} = (0.659 \pm 0.015)[Fe/H]_{\rm LASP} - (0.081 \pm 0.003)\ (\rm [Fe/H]\ \epsilon\ [-0.5, 0.1]) \\  
\rm [Fe/H]_{\rm KIC} = (0.842 \pm 0.055)[Fe/H]_{\rm LASP} - (0.083 \pm 0.012)\ (\rm [Fe/H]\ \epsilon\ [0.1, 0.4]) \: ,
           \end{array} \right. \: 
\end{eqnarray}

\noindent
and dwarfs: 
\begin{eqnarray} 
 \left\{ \begin{array}{l}
\rm [Fe/H]_{\rm KIC} = (0.400 \pm 0.010)[Fe/H]_{\rm LASP} - (0.206 \pm 0.002)\ (\rm [Fe/H]\ \epsilon\ [-0.5, 0.2]) \\  
\rm [Fe/H]_{\rm KIC} = (0.651 \pm 0.055)[Fe/H]_{\rm LASP} - (0.228 \pm 0.017)\ (\rm [Fe/H]\ \epsilon\ [0.2, 0.5]) ,
 \end{array} \right.
\end{eqnarray}

\noindent
to calibrate the parameters in the KIC.

\section{Conclusions and Prospects}
\label{sect:8}
The low-resolution spectroscopic observations for the stars in the \kepler\ field with \lamost\ were started at Xinglong observatory on May 30, 2011.  There are 88,628 flux- and wavelength-calibrated, sky-subtracted spectra to be released in the third data release (DR3) of \lamost. These spectra were obtained in the observation seasons from June 2012 to September 2014. After three years of discontinuous observations, 14 LK-fields covering the whole \kepler-field have been observed discontinuously at 25 nights. Based on the \lasp\ stellar parameters, we derived the analysis stellar atmospheric parameters (\teff, \logg, \feh), the radial velocity (\vrad) and their errors for 51,406 stars from 61,226 late A, F, G and K type stars having at least one \lamost\ spectrum with \snrg\ $\geq$ 6.0 in the catalog of the \project. The magnitude distribution of these objects mainly ranges from 11 to 15 magnitude. Among the 51,406 observed stars, more than half (30,110) were observed photometrically by the Kepler mission in the \kepler\ field. As 8,632 targets have been observed more than once, the ratio of multiple observation stars is close to 17\%. The wrong IDs of 9 fibers of spectrograph 4 affect 71 spectra in the \project\ since the start of the survey in June 2012. We have corrected the ID of these spectra according to the latest updated information in the \lamost\ DR2 website. 

The comparison of the \lasp\ parameters with the \teff\ and \feh\ derived from high-resolution (R $\geq$ 20,000) spectra and the \logg\ calculated by the means of asteroseismology given in six subsamples in HT4 offers an effective method to perform the external calibration of the parameters from the \project. Using the atmospheric parameters and their errors of the common giant and dwarf stars in the two data sets, close to 1:1 linear relations are found as best fits for \teff, \logg, and \feh, respectively. Almost all data points are located in the 3$\sigma$ confidence intervals of the residual parameters. The mean deviation (\lasp\ - Huber) and the external uncertainties of \teff, \logg\ and \feh\ are 34$\pm$131 K, 0.13$\pm$0.19 dex and  -0.04$\pm$0.15 dex for the giants, 5$\pm$104 K, -0.05$\pm$0.16 dex and -0.01$\pm$0.10 dex for the dwarfs, respectively. We showed that the \lasp\ parameters of the stars in our catalogue are reliable in the ranges of 3,800 to 5,300 K for \teff,  1.5 to 3.5 dex for \logg, and -1.0 to 0.5 dex for \feh\ for giant stars, 4,000 to 7,000 K for \teff, 3.6 to 4.9 dex for \logg, and -0.6 to 0.5 dex for \feh\ for dwarf stars, respectively. These ranges are not the same for giant and dwarf stars, as shown in the pink background regions of Figure \ref{fig9}. Using the unbiased estimated method, we estimated the internal errors of the \lasp\ parameters with \snrg\ $\geq$ 6.0 based on the stars with multiple \lamost\ spectra: 91 K, 0.12 dex and 0.09 dex for \teff, \logg\ and \feh, respectively. The internal uncertainties of stellar parameters are defined with second-order polynomials as a function of \snrg. Finally, the \lasp\ stellar parameters and their errors were calibrated by taking both internal and external calibration into account.

For the analysis of the overall distributions of the calibrated atmospheric parameters and radial velocities to search for particular stars, we considered the \lasp\ parameters derived from the \lamost\ spectra with the highest \snrg\ value for stars with multiple observations. No special objects were found from the analysis of the effective temperature and the surface gravity. However, 
106 stars could be classified as candidate MPs and 9 as candidate VMPs from the sample of stars with \feh~$<$ -1.0 dex after visual inspection of the results. Moreover, 18 stars from the sample of targets with \vrad\ $<$ -300~km/s in \kepler\ field can be considered as candidate HVs.  

There are 51,399 common stars between the \lasp\ catalogue and the KIC. For 41,775 of them, atmospheric parameters are listed in the KIC. Issues with the KIC parameters, especially for \logg\ and \feh, are confirmed by the comparison with the calibrated parameters derived within the \project. The KIC overestimates the parameters for most of giants but underestimates the \feh\ values in the range of \feh\ $>$ -0.2 dex. All the parameters in the KIC can be calibrated by applying the correlations between \lasp\ and KIC values as given in Section \ref{sect:7}. The missing atmospheric parameters for 9,624 stars in the KIC can now be replaced for the majority of them by the calibrated \lasp\ parameters within the reliable ranges (3,800 $\sim$ 7,000 K for \teff, 1.5 $\sim$ 4.9 dex for \logg, and -1.0 $\sim$ 0.5 dex for \feh).

The stellar parameters determined by the \lasp\ based on the normalized low-resolution spectra of the \project\ are available for \lamost\ users from on the world wide web: \href{url}{http://dr3.lamost.org}. The reduced spectra and stellar parameters will be released to the interested astronomers all over the world through the \lamost\ official data release in June 2017. The calibrated stellar atmospheric parameters and the radial velocities obtained within the \project\ are helpful for various studies of stars in the \kepler\ field.  Although plenty of the stars that have been observed by the \kepler\ mission were not observed yet during the first round of observations, a good progress has been made in 2015 as 32 additional plates, covering the whole \kepler\ field except for one subfield (LK01), have been observed during 18 nights. The observations for the \project\ are still ongoing. Moreover, a proposal for \lamost\ observations for six K2 mission fields with the declination higher than -10 deg \citep{How2014} has been approved and started to carry out. We therefore expect to obtain many more high-quality spectroscopic observations with the \lamost\ for stars in both the \kepler\ field and fields of the K2 mission in near future.

\begin{deluxetable}{llrclrrrl}
\tablewidth{0pt}
\tabletypesize{\scriptsize}
\tablecaption{The calibrated \lasp\ atmosphere parameters (\teff, \logg\ and \feh) and the \lasp\ radial velocity (\vrad) for the candidates of the particular objects in the \project.  The spectra of these candidates were inspected visually to ensure the reliability of these candidates. \label{Tab5}}
\tablehead{ \colhead{Obsid} & \colhead{Target} & \colhead{\snrg} & \colhead{Subclass} & \colhead{\teff\ (K)} & \colhead{\logg\ (dex)} &  \\ 
}
\startdata
243016087  &  KIC06022596  &   57.54  &  F2    & $5176\pm108$  &  $2.421\pm0.153$  &  $-1.728\pm0.120$  &  $-130.59\pm39.95$  & MPs\\
243116087  &  KIC06022596  &   56.15  &  F2    & $5208\pm108$  &  $2.480\pm0.154$  &  $-1.702\pm0.121$  &  $-128.07\pm38.60$  & MPs\\
163007106  &  KIC06022596  &   46.66  &  F2    & $5221\pm110$  &  $2.496\pm0.157$  &  $-1.719\pm0.123$  &  $-134.74\pm31.90$  & MPs\\
238003154  &  KIC09933034  &  109.00  &  F0    & $6232\pm86 $  &  $4.255\pm0.126$  &  $-1.143\pm0.082$  &  $-123.38\pm39.38$  & MPs\\
242703154  &  KIC09933034  &   81.51  &  F0    & $6262\pm88 $  &  $4.289\pm0.129$  &  $-1.132\pm0.085$  &  $-124.43\pm41.80$  & MPs\\
242603154  &  KIC09933034  &   18.03  &  F0    & $6226\pm114$  &  $4.192\pm0.172$  &  $-1.173\pm0.121$  &  $-123.70\pm42.52$  & MPs\\
154015047  &  KIC08162514  &   66.86  &  G5    & $4431\pm106$  &  $1.106\pm0.151$  &  $-1.289\pm0.118$  &  $ -73.82\pm17.77$  & MPs\\
155015047  &  KIC08162514  &   49.99  &  G8    & $4474\pm109$  &  $1.198\pm0.156$  &  $-1.225\pm0.122$  &  $ -77.47\pm18.08$  & MPs\\
159612061  &  KIC08364751  &  234.41  &  F0    & $6244\pm83 $  &  $4.232\pm0.120$  &  $-1.098\pm0.076$  &  $-296.47\pm35.32$  & MPs\\
247610021  &  KIC08364751  &   79.37  &  F3    & $6192\pm88 $  &  $4.257\pm0.130$  &  $-1.237\pm0.085$  &  $-304.03\pm38.02$  & MPs\\
170509015  &  KIC09432243  &  113.34  &  F0    & $5867\pm86 $  &  $4.030\pm0.125$  &  $-1.944\pm0.081$  &  $-161.10\pm50.09$  & MPs\\
165909015  &  KIC09432243  &   51.26  &  F0    & $5884\pm93 $  &  $4.057\pm0.138$  &  $-1.953\pm0.093$  &  $-158.96\pm47.14$  & MPs\\
238004170  &  KIC09813342  &  165.85  &  G0    & $5657\pm84 $  &  $4.315\pm0.122$  &  $-1.328\pm0.078$  &  $-172.55\pm29.48$  & MPs\\
242604170  &  KIC09813342  &  116.47  &  G0    & $5648\pm86 $  &  $4.308\pm0.125$  &  $-1.339\pm0.081$  &  $-171.77\pm31.47$  & MPs\\
249201249  &  KIC09818964  &   73.31  &  G2    & $4923\pm105$  &  $2.085\pm0.150$  &  $-1.662\pm0.117$  &  $-200.73\pm28.13$  & MPs\\
242706043  &  KIC09818964  &   45.82  &  G3    & $4880\pm110$  &  $1.783\pm0.158$  &  $-1.702\pm0.124$  &  $-205.47\pm28.89$  & MPs\\
247611205  &  KIC09956941  &   73.33  &  F7    & $5907\pm89 $  &  $4.068\pm0.131$  &  $-1.849\pm0.087$  &  $-285.20\pm37.92$  & MPs\\
241006143  &  KIC09956941  &   48.83  &  F0    & $5890\pm94 $  &  $4.055\pm0.139$  &  $-1.855\pm0.094$  &  $-284.24\pm43.81$  & MPs\\
238015073  &  KIC10319045  &  215.86  &  G5    & $4432\pm100$  &  $1.228\pm0.141$  &  $-1.077\pm0.110$  &  $-126.78\pm16.73$  & MPs\\
242615073  &  KIC10319045  &  150.78  &  G5    & $4430\pm101$  &  $1.207\pm0.142$  &  $-1.071\pm0.111$  &  $-128.31\pm16.78$  & MPs\\
242612192  &  KIC11017176  &   90.15  &  G7    & $4847\pm104$  &  $1.506\pm0.147$  &  $-1.474\pm0.115$  &  $-226.61\pm24.90$  & MPs\\
238012192  &  KIC11017176  &   68.40  &  G7    & $4914\pm106$  &  $1.660\pm0.151$  &  $-1.376\pm0.118$  &  $-229.65\pm25.02$  & MPs\\
157214098  &  KIC11457596  &   58.31  &  F0    & $6066\pm92 $  &  $4.158\pm0.135$  &  $-1.492\pm0.090$  &  $-285.17\pm47.70$  & MPs\\
248606006  &  KIC11457596  &   47.75  &  F0    & $6300\pm94 $  &  $4.273\pm0.139$  &  $-1.377\pm0.094$  &  $-286.85\pm37.38$  & MPs\\
248610109  &  KIC11704816  &   59.59  &  G2    & $5089\pm107$  &  $2.416\pm0.153$  &  $-1.624\pm0.120$  &  $-180.04\pm25.69$  & MPs\\
249610109  &  KIC11704816  &   17.24  &  G2    & $5162\pm128$  &  $2.903\pm0.188$  &  $-1.569\pm0.146$  &  $-183.21\pm22.67$  & MPs\\
249209055  &  KIC11857234  &   52.26  &  G3    & $4749\pm109$  &  $1.683\pm0.155$  &  $-1.370\pm0.122$  &  $-108.28\pm17.06$  & MPs\\
248610023  &  KIC11857234  &   35.12  &  G3    & $4680\pm114$  &  $1.624\pm0.164$  &  $-1.445\pm0.128$  &  $-110.76\pm20.65$  & MPs\\
52507223   &  KIC01580348  &   32.90  &  G3    & $4635\pm115$  &  $1.122\pm0.166$  &  $-1.670\pm0.130$  &  $-316.10\pm22.83$  & MPs\\
243006015  &  KIC03838579  &   59.70  &  F2    & $5928\pm91 $  &  $4.204\pm0.134$  &  $-1.036\pm0.090$  &  $ -93.67\pm28.97$  & MPs\\
243104137  &  KIC04446192  &  132.63  &  F7    & $5827\pm85 $  &  $4.303\pm0.124$  &  $-1.240\pm0.080$  &  $ -70.27\pm36.89$  & MPs\\
243013134  &  KIC05176287  &   16.23  &  K1    & $4597\pm130$  &  $2.392\pm0.191$  &  $-1.074\pm0.148$  &  $ -45.26\pm17.52$  & MPs\\
159601033  &  KIC05268275  &  102.09  &  F2    & $5791\pm86 $  &  $4.357\pm0.126$  &  $-1.049\pm0.082$  &  $  -7.73\pm31.64$  & MPs\\
154001194  &  KIC05271670  &   32.19  &  G2    & $5537\pm101$  &  $4.453\pm0.150$  &  $-1.047\pm0.103$  &  $  -7.77\pm27.57$  & MPs\\
154007230  &  KIC05362244  &   46.72  &  F5    & $6017\pm95 $  &  $4.253\pm0.140$  &  $-1.062\pm0.095$  &  $-104.87\pm32.72$  & MPs\\
163002116  &  KIC05422924  &   31.15  &  G2    & $5191\pm116$  &  $2.897\pm0.168$  &  $-1.390\pm0.131$  &  $ -77.31\pm32.11$  & MPs\\
243114093  &  KIC05513197  &   37.08  &  G2    & $5115\pm113$  &  $2.303\pm0.163$  &  $-1.138\pm0.127$  &  $ -83.83\pm30.56$  & MPs\\
163002084  &  KIC05933607  &   44.88  &  G3    & $4962\pm111$  &  $2.282\pm0.158$  &  $-1.045\pm0.124$  &  $-210.33\pm24.51$  & MPs\\
243112023  &  KIC05948716  &   49.52  &  F0    & $6308\pm94 $  &  $4.129\pm0.138$  &  $-1.382\pm0.093$  &  $-384.46\pm44.13$  & MPs\\
52815099   &  KIC05966097  &   26.00  &  F3    & $6142\pm105$  &  $4.204\pm0.157$  &  $-1.505\pm0.109$  &  $-373.05\pm40.29$  & MPs\\
154002171  &  KIC06111652  &   63.90  &  G5    & $4674\pm107$  &  $1.715\pm0.152$  &  $-1.024\pm0.119$  &  $ -16.79\pm20.34$  & MPs\\
163005015  &  KIC06263778  &   42.56  &  G3    & $4902\pm111$  &  $2.179\pm0.159$  &  $-1.319\pm0.125$  &  $-292.35\pm25.34$  & MPs\\
161401216  &  KIC06271226  &   56.32  &  G6    & $4721\pm108$  &  $1.326\pm0.154$  &  $-1.372\pm0.121$  &  $ -65.14\pm21.75$  & MPs\\
162314060  &  KIC06533598  &   11.14  &  G7    & $4639\pm129$  &  $4.331\pm0.196$  &  $-1.187\pm0.140$  &  $ -19.61\pm21.82$  & MPs\\
159608008  &  KIC06604237  &  130.46  &  G3    & $4646\pm102$  &  $1.307\pm0.143$  &  $-1.884\pm0.112$  &  $-330.28\pm30.71$  & MPs\\
159610225  &  KIC06766131  &   53.09  &  F2    & $5793\pm93 $  &  $3.883\pm0.137$  &  $-1.013\pm0.092$  &  $-104.06\pm38.28$  & MPs\\
159603018  &  KIC06769852  &   60.43  &  G0    & $5432\pm107$  &  $3.272\pm0.153$  &  $-1.122\pm0.120$  &  $-222.36\pm27.77$  & MPs\\
161402195  &  KIC07263702  &   32.33  &  F6    & $5782\pm101$  &  $4.273\pm0.150$  &  $-1.199\pm0.103$  &  $ -22.24\pm40.14$  & MPs\\
163004071  &  KIC07505345  &   43.02  &  F0    & $6664\pm96 $  &  $4.251\pm0.142$  &  $-1.010\pm0.096$  &  $-235.09\pm34.11$  & MPs\\
161410216  &  KIC07590793  &   33.56  &  G6    & $5156\pm100$  &  $4.533\pm0.149$  &  $-1.268\pm0.102$  &  $-260.26\pm21.75$  & MPs\\
247605091  &  KIC07614421  &   26.60  &  G3    & $4969\pm119$  &  $2.135\pm0.172$  &  $-1.660\pm0.134$  &  $-243.66\pm27.27$  & MPs\\
161410078  &  KIC07665025  &   37.05  &  G3    & $5024\pm113$  &  $2.419\pm0.163$  &  $-1.301\pm0.127$  &  $-277.68\pm27.26$  & MPs\\
170501034  &  KIC07917764  &   59.51  &  G3    & $4829\pm107$  &  $1.670\pm0.153$  &  $-1.256\pm0.120$  &  $ -34.16\pm25.81$  & MPs\\
163015114  &  KIC07940280  &   67.72  &  F0    & $6016\pm90 $  &  $4.192\pm0.132$  &  $-1.302\pm0.088$  &  $-293.76\pm37.00$  & MPs\\
161404092  &  KIC07948268  &   90.16  &  G2    & $5154\pm104$  &  $2.870\pm0.147$  &  $-1.196\pm0.115$  &  $-296.78\pm26.41$  & MPs\\
161408041  &  KIC08019664  &   26.97  &  G3    & $4993\pm119$  &  $2.287\pm0.172$  &  $-1.433\pm0.134$  &  $-392.03\pm29.72$  & MPs\\
163015138  &  KIC08077380  &   72.41  &  G2    & $5373\pm89 $  &  $3.565\pm0.131$  &  $-1.090\pm0.087$  &  $-179.12\pm25.33$  & MPs\\
163013110  &  KIC08082012  &   33.85  &  G5    & $4481\pm115$  &  $1.100\pm0.165$  &  $-1.175\pm0.129$  &  $-203.86\pm17.00$  & MPs\\
247603025  &  KIC08237832  &   50.14  &  F6    & $5919\pm94 $  &  $3.658\pm0.138$  &  $-1.126\pm0.093$  &  $-233.69\pm24.69$  & MPs\\
163016116  &  KIC08409682  &   34.82  &  G8    & $5079\pm100$  &  $4.786\pm0.148$  &  $-1.363\pm0.101$  &  $ -45.90\pm19.19$  & MPs\\
163012135  &  KIC08412954  &   39.96  &  G2    & $5129\pm112$  &  $2.594\pm0.161$  &  $-1.939\pm0.126$  &  $-285.53\pm37.45$  & MPs\\
163011233  &  KIC08476245  &   24.69  &  G3    & $4977\pm120$  &  $2.059\pm0.175$  &  $-1.035\pm0.136$  &  $-124.18\pm20.85$  & MPs\\
161415241  &  KIC08612146  &   29.59  &  G3    & $4857\pm117$  &  $1.778\pm0.169$  &  $-1.821\pm0.132$  &  $-196.22\pm29.30$  & MPs\\
242607068  &  KIC08869235  &   47.17  &  G6    & $5014\pm110$  &  $2.607\pm0.157$  &  $-1.020\pm0.123$  &  $-217.90\pm20.56$  & MPs\\
242602239  &  KIC08999218  &   59.87  &  F0    & $6114\pm91 $  &  $4.218\pm0.134$  &  $-1.153\pm0.090$  &  $ -64.38\pm40.37$  & MPs\\
242707109  &  KIC09004948  &   58.37  &  G6    & $4954\pm108$  &  $2.115\pm0.153$  &  $-1.718\pm0.120$  &  $-252.13\pm29.68$  & MPs\\
238005009  &  KIC09071237  &  105.14  &  G5    & $5293\pm86 $  &  $4.552\pm0.126$  &  $-1.211\pm0.082$  &  $-264.42\pm21.28$  & MPs\\
241007071  &  KIC09156667  &   43.72  &  F2    & $5515\pm96 $  &  $3.756\pm0.141$  &  $-1.629\pm0.096$  &  $-265.32\pm35.29$  & MPs\\
158804247  &  KIC09245734  &   90.99  &  F5    & $5874\pm87 $  &  $3.971\pm0.128$  &  $-1.580\pm0.084$  &  $ -44.44\pm41.96$  & MPs\\
158814069  &  KIC09358384  &   48.54  &  A6IV  & $6568\pm94 $  &  $4.368\pm0.139$  &  $-1.218\pm0.094$  &  $   3.45\pm40.35$  & MPs\\
238006155  &  KIC09452906  &   26.52  &  F5    & $6062\pm105$  &  $4.173\pm0.156$  &  $-1.177\pm0.109$  &  $ -15.66\pm35.28$  & MPs\\
158804038  &  KIC09610507  &   33.59  &  G2    & $5648\pm100$  &  $3.802\pm0.149$  &  $-1.001\pm0.102$  &  $  -6.84\pm30.35$  & MPs\\
242606141  &  KIC09637337  &  129.71  &  G3    & $5055\pm102$  &  $2.681\pm0.143$  &  $-1.007\pm0.112$  &  $-143.79\pm22.08$  & MPs\\
242608054  &  KIC09696716  &  135.81  &  G3    & $4944\pm102$  &  $2.088\pm0.143$  &  $-1.479\pm0.112$  &  $-154.07\pm28.61$  & MPs\\
247711176  &  KIC09836233  &   33.43  &  K3    & $4836\pm100$  &  $4.768\pm0.149$  &  $-1.035\pm0.102$  &  $-141.19\pm12.74$  & MPs\\
249207177  &  KIC10398120  &  162.33  &  G5    & $4686\pm101$  &  $1.620\pm0.142$  &  $-1.002\pm0.111$  &  $-206.11\pm13.10$  & MPs\\
242613111  &  KIC10521392  &   68.65  &  F5    & $6002\pm90 $  &  $4.179\pm0.132$  &  $-1.059\pm0.088$  &  $ -13.37\pm33.87$  & MPs\\
249601024  &  KIC10729186  &   22.96  &  G7    & $5156\pm122$  &  $3.161\pm0.177$  &  $-1.087\pm0.138$  &  $ -83.42\pm18.34$  & MPs\\
154803010  &  KIC10737052  &   35.70  &  G4    & $4954\pm114$  &  $2.233\pm0.164$  &  $-1.266\pm0.128$  &  $-252.20\pm25.11$  & MPs\\
249601045  &  KIC10858420  &   11.44  &  G7    & $5281\pm128$  &  $4.023\pm0.194$  &  $-1.089\pm0.139$  &  $-119.22\pm15.94$  & MPs\\
241009055  &  KIC10920437  &   71.12  &  G2    & $5241\pm106$  &  $2.839\pm0.150$  &  $-1.271\pm0.118$  &  $-159.40\pm26.72$  & MPs\\
157213090  &  KIC11044756  &   27.47  &  F6    & $5645\pm104$  &  $3.900\pm0.155$  &  $-1.227\pm0.108$  &  $-242.86\pm28.43$  & MPs\\
157214239  &  KIC11296574  &  109.35  &  G0    & $5580\pm86 $  &  $3.848\pm0.125$  &  $-1.160\pm0.081$  &  $-260.01\pm34.63$  & MPs\\
249213204  &  KIC11345077  &   37.45  &  G3    & $5104\pm113$  &  $2.726\pm0.163$  &  $-1.255\pm0.127$  &  $-245.21\pm20.36$  & MPs\\
250006172  &  KIC11563791  &  129.19  &  G6    & $4947\pm102$  &  $2.446\pm0.143$  &  $-1.092\pm0.112$  &  $-269.72\pm19.00$  & MPs\\
248603228  &  KIC11757807  &  117.74  &  F9    & $4659\pm102$  &  $1.522\pm0.144$  &  $-1.388\pm0.113$  &  $-220.34\pm18.30$  & MPs\\
249215021  &  KIC11855373  &   80.07  &  A6IV  & $6648\pm88 $  &  $4.362\pm0.129$  &  $-1.466\pm0.085$  &  $-243.65\pm26.38$  & MPs\\
249215154  &  KIC11953764  &   75.52  &  G3    & $5169\pm105$  &  $3.010\pm0.149$  &  $-1.053\pm0.117$  &  $-177.25\pm19.89$  & MPs\\
249211077  &  KIC12004528  &   76.55  &  F0    & $6202\pm89 $  &  $4.117\pm0.130$  &  $-1.785\pm0.086$  &  $-199.49\pm41.33$  & MPs\\
249613052  &  KIC12017985  &   53.65  &  G3    & $4941\pm109$  &  $2.143\pm0.155$  &  $-1.754\pm0.121$  &  $-200.09\pm26.24$  & MPs\\
249216059  &  KIC12051330  &   86.01  &  G3    & $4916\pm104$  &  $2.120\pm0.147$  &  $-1.435\pm0.116$  &  $ -69.32\pm20.47$  & MPs\\
248604206  &  KIC12207740  &  137.58  &  G6    & $4948\pm102$  &  $2.361\pm0.143$  &  $-1.127\pm0.112$  &  $ -11.63\pm20.50$  & MPs\\
249609181  &  KIC12210298  &   38.81  &  F0    & $6413\pm98 $  &  $4.261\pm0.145$  &  $-1.342\pm0.099$  &  $-169.33\pm31.90$  & MPs\\
157211215  &  KIC12216301  &   63.31  &  G2    & $5220\pm107$  &  $2.937\pm0.152$  &  $-1.341\pm0.119$  &  $-217.26\pm29.95$  & MPs\\
248615245  &  KIC12304604  &   39.25  &  G0    & $5251\pm113$  &  $2.985\pm0.161$  &  $-1.305\pm0.126$  &  $-182.56\pm23.70$  & MPs\\
248615168  &  KIC12645981  &   88.89  &  F0    & $6514\pm87 $  &  $4.224\pm0.128$  &  $-1.003\pm0.084$  &  $-294.27\pm28.49$  & MPs\\
238014090  &  KIC10383102  &   36.47  &  F0    & $5957\pm99 $  &  $3.676\pm0.146$  &  $-2.257\pm0.100$  &  $-150.17\pm53.56$  & VMPs\\
242714090  &  KIC10383102  &   32.87  &  F0    & $5846\pm101$  &  $3.704\pm0.149$  &  $-2.192\pm0.103$  &  $-153.54\pm56.68$  & VMPs\\
247607120  &  KIC07693833  &  113.55  &  G2    & $4931\pm102$  &  $1.982\pm0.145$  &  $-2.292\pm0.113$  &  $ -13.08\pm27.59$  & VMPs\\
242707138  &  KIC09006890  &   41.86  &  G2    & $4823\pm112$  &  $1.829\pm0.160$  &  $-2.236\pm0.125$  &  $-270.85\pm43.57$  & VMPs\\
248602055  &  KIC11080134  &   66.36  &  F0    & $5822\pm90 $  &  $3.626\pm0.133$  &  $-2.133\pm0.088$  &  $-266.92\pm42.09$  & VMPs\\
------------\\
242603099  &  KIC09751081  &   85.62  &  G7    & $4782\pm104$  &  $1.680\pm0.148$  &  $-1.697\pm0.116$  &  $-311.77\pm27.01$  & HVs\\
242703076  &  KIC09751081  &   53.67  &  G7    & $4778\pm109$  &  $1.674\pm0.155$  &  $-1.719\pm0.121$  &  $-309.25\pm28.22$  & HVs\\
154007087  &  KIC05621880  &   34.48  &  G3    & $5582\pm100$  &  $4.239\pm0.148$  &  $-0.654\pm0.102$  &  $-316.05\pm20.27$  & HVs\\
159603161  &  KIC07191496  &  178.88  &  G6    & $4900\pm101$  &  $1.968\pm0.141$  &  $-1.983\pm0.111$  &  $-305.52\pm33.37$  & HVs\\
163014201  &  KIC07658030  &   26.07  &  G5    & $4303\pm119$  &  $0.415\pm0.173$  &  $-1.905\pm0.135$  &  $-406.20\pm23.55$  & HVs\\
242707005  &  KIC08802291  &   27.52  &  G8    & $4707\pm118$  &  $1.495\pm0.171$  &  $-1.775\pm0.134$  &  $-308.53\pm27.56$  & HVs\\
250007083  &  KIC11032723  &   31.34  &  G6    & $4606\pm116$  &  $1.552\pm0.167$  &  $-1.251\pm0.131$  &  $-304.73\pm13.96$  & HVs\\
249209210  &  KIC11395462  &   66.91  &  G6    & $4764\pm106$  &  $2.194\pm0.151$  &  $-0.764\pm0.118$  &  $-328.13\pm13.78$  & HVs\\
154811189  &  KIC12019793  &   20.46  &  G6    & $4888\pm124$  &  $2.543\pm0.181$  &  $-0.655\pm0.141$  &  $-303.65\pm18.96$  & HVs\\
-----------\\
243009061  &  KIC05691816  &   52.75  &  G7    & $4957\pm109$  &  $2.071\pm0.155$  &  $-1.638\pm0.121$  &  $-305.47\pm31.85$  & MPs,HVs\\
154005087  &  KIC06116549  &   68.50  &  G2    & $4993\pm106$  &  $2.260\pm0.151$  &  $-1.799\pm0.118$  &  $-359.86\pm40.14$  & MPs,HVs\\
154008102  &  KIC06362206  &  142.61  &  F0    & $5995\pm84 $  &  $4.052\pm0.123$  &  $-1.441\pm0.079$  &  $-324.81\pm39.04$  & MPs,HVs\\
154004124  &  KIC07030715  &  103.78  &  F0    & $6293\pm86 $  &  $4.194\pm0.126$  &  $-1.332\pm0.082$  &  $-336.47\pm39.84$  & MPs,HVs\\
161405185  &  KIC07666893  &   35.28  &  G3    & $4542\pm114$  &  $0.884\pm0.164$  &  $-1.880\pm0.128$  &  $-316.54\pm25.42$  & MPs,HVs\\
154014085  &  KIC07673401  &   42.07  &  F0    & $5914\pm96 $  &  $3.989\pm0.142$  &  $-1.699\pm0.097$  &  $-305.13\pm53.01$  & MPs,HVs\\
161409033  &  KIC08680868  &   27.97  &  G4    & $4926\pm118$  &  $2.346\pm0.171$  &  $-1.019\pm0.133$  &  $-381.01\pm20.66$  & MPs,HVs\\
241002039  &  KIC09335536  &   70.25  &  G6    & $4819\pm106$  &  $1.670\pm0.150$  &  $-1.576\pm0.118$  &  $-356.46\pm27.98$  & MPs,HVs\\
241004161  &  KIC10203516  &  119.67  &  F5    & $6026\pm85 $  &  $4.191\pm0.125$  &  $-1.235\pm0.081$  &  $-305.50\pm38.29$  & MPs,HVs\\
249214199  &  KIC11336325  &   63.02  &  G5    & $4374\pm107$  &  $0.936\pm0.152$  &  $-1.250\pm0.119$  &  $-326.35\pm14.06$  & MPs,HVs\\

\enddata
\end{deluxetable}

\section{Acknowledgments}
Guoshoujing Telescope (the Large Sky Area Multi-Object Fiber Spectroscopic Telescope \lamost) is a National Major Scientific Project built by the Chinese Academy of Sciences. Funding for the project has been provided by the National Development and Reform Commission. \lamost\ is operated and managed by the National Astronomical Observatories, Chinese Academy of Sciences. JNF acknowledges the support by the Joint Fund of Astronomy of National Science Foundation of China (NSFC) and Chinese Academy of Sciences through the grant U1231202 and the National Basic Research Program of China (973 program 2014CB845700 and 2013CB834900). The LAMOST FELLOWSHIP is supported by Special Funding for Advanced Users, budgeted and administrated by Center for Astronomical Mega-Science, Chinese Academy of Sciences (CAMS). YW acknowledges the NSFC under grant 11403056. SD is supported by ‘the Strategic Priority Research Program-The Emergence of Cosmological Structures’ of the Chinese Academy of Sciences (Grant No. XDB09000000) and Project 11573003 supported by NSFC. ABR thanks Antonio Frasca, Joanna Molenda-$\dot{\rm Z}$akowicz and Gianni Catanzaro for helpful discussions.

\end{document}